\documentclass[12pt,preprint]{aastex}

\shorttitle{Exoplanet Host Star Diameters}
\shortauthors{Baines et al.}

\begin{document}

\title{CHARA Array Measurements of the Angular Diameters of Exoplanet Host Stars}

\author{Ellyn K. Baines, Harold A. McAlister, Theo A. ten Brummelaar, Nils~H.~Turner, Judit Sturmann, Laszlo Sturmann, \& P.~J. Goldfinger}
\affil{Center for High Angular Resolution Astronomy, Georgia State University, P.O. Box 3969, Atlanta, GA 30302-3969}
\email{baines, hal@chara.gsu.edu; theo, nils, judit, sturmann, pj@chara-array.org}

\author{Stephen T. Ridgway}
\affil{Kitt Peak National Observatory, National Optical Astronomy Observatory, P.O. Box 26732, Tucson, AZ 85726-6732} 
\email{ridgway@noao.edu}

\altaffiltext{}{For preprints, please email baines@chara.gsu.edu.}

\begin{abstract}
We have measured the angular diameters for a sample of 24 exoplanet host stars using Georgia State University's CHARA Array interferometer. We use these improved angular diameters together with \emph{Hipparcos} parallax measurements to derive linear radii and to estimate the stars' evolutionary states.
\end{abstract}


\section{Introduction}
Nearly 300 exoplanet systems are now known, discovered via radial velocity surveys and photometric transit events. Most known exoplanet host stars are Sun-like in nature, and their planets have minimum masses comparable to Saturn with orbital semimajor axes ranging from 0.04 to 6.0~AU \citep{2005ApJ...619..570M}, painting pictures of planetary systems very different from our own.

Many exoplanet host stars' angular diameters have been estimated using photometric or spectroscopic methods. For example, \citet{2003A&A...411L.501R} matched 2MASS infrared photometry to synthetic photometry in order to estimate stellar temperatures, which then produced angular diameter estimations. \citet{2005ApJ...622.1102F} performed the first uniform spectroscopic analysis for all the exoplanets' host stars known at the time as well as for a large sample of single stars. They determined the effective temperature $T_{\rm eff}$, log~$g$, $v \sin i$, and metallicity for 1040 FGK-type stars to check if there were any correlations between stellar metallicity and the presence of planets and found a rapid rise in the fraction of stars with planets for high-metallicity stars. They calculated stellar radii using stellar luminosities derived from $T_{\rm eff}$, \emph{Hipparcos} parallaxes, and a bolometric correction.

These methods are useful for estimating stellar sizes, but are inherently indirect in nature. Interferometric observations directly measure the angular diameters for these stars, which, in conjunction with parallaxes, lead to linear radii.

\section{Interferometric Observations}
The target list was derived from the general exoplanet list using declination limits (north of -10$^\circ$) and magnitude constraints. The stars needed to be brighter than $V=+10$ in order for the tip/tilt subsystem to lock onto the star and brighter than $K=+6.5$ so fringes were easily visible. This reduced the exoplanet list to approximately 80 targets, and we obtained data on 24 of them over multiple observing runs spanning 2004 January to 2007 September (see~Table \ref{observations}).

The stars were observed using the Center for High Angular Resolution Astronomy (CHARA) Array, a six-element Y-shaped interferometric array located on Mount Wilson, California \citep{2005ApJ...628..453T}. The Array presently employs visible wavelengths (470-800 nm) for tracking and tip/tilt corrections and near infrared bands ($H$ at 1.67~$\mu$m and $K'$ at 2.15~$\mu$m) for fringe detection and data collection. All observations were obtained using the pupil-plane ``CHARA Classic'' beam combiner in the $K'$-band, except observations of HD~189733, which were obtained using the $H$-band. The observing procedure and data reduction process employed here are described in \citet{2005ApJ...628..439M}.

Most observations were taken using the longest baseline the CHARA Array offers, 331~m for the S1-E1 pair of telescopes, due to its sensitivity in measuring stellar diameters\footnote{The three arms of the Array are denoted by their cardinal directions: ``S'' is south, ``E'' is east, and ``W'' is west. Each arm bears two telescopes, numbered ``1'' for the telescope farthest from the beam combining laboratory and ``2'' for the telescope closer to the lab.}. If no S1-E1 data were obtained, the observations with the longest available baseline were used in the angular diameter measurement. Table \ref{observations} lists the exoplanet host stars observed, their calibrators, the baseline used, the dates of the observations, and the number of observations obtained. More information on the transiting planet system HD~189733 can be found in \citet{2007ApJ...661L.195B}.

We used the standard calibrator-target-calibrator observing pattern so that every target was flanked by calibrator observations made as close in time as possible. This allowed us to calculate the target's calibrated visibilities from the raw visibilities of the target and calibrator. Acceptable calibrators were chosen to have expected visibility amplitudes greater than 85\% on the baselines used, and the high visibilities meant the calibrators were nearly unresolved. Therefore, uncertainties in the calibrator's diameter do not affect the target's diameter calculation as much as if the calibrator star had a significant angular size on the sky.

Another small source of potential systematic error in the target's diameter measurement arises from limb-darkening effects, though the error is on the order of a few percent and is not as significant an effect in the $K'$-band as it would be for measurements in visible wavelengths \citep{2006ApJ...644..475B}. For barely resolved calibrators, this effect is negligible.

In an effort to find reliable calibrators, we made spectral energy distribution (SED) fits based on published $UBVRIJHK$ photometric values for each calibrator to establish diameter estimates and to check if there was any excess emission associated with a low-mass stellar companion or circumstellar disk. Calibrator candidates with variable radial velocities reported in the literature or with any other indication of a possible companion were discarded even if their SEDs displayed no characteristics of duplicity.

Limb-darkened angular diameter estimates for the calibrators were determined using Kurucz model atmospheres\footnote{Available to download at http://kurucz.cfa.harvard.edu.} based on $T_{\rm eff}$ and log~$g$ values obtained from the literature. The models were then fit to observed photometric values also from the literature after converting magnitudes to fluxes using \citet{1996AJ....112..307C} for $UBVRI$ values and \citet{2003AJ....126.1090C} for $JHK$ values. See Table \ref{calibrators} for the $T_{\rm eff}$ and log~$g$ used and the resulting limb-darkened angular diameters.

Table~\ref{exp_calib_visy} lists the Modified Julian Date (MJD), baseline ($B$), position angle ($\Theta$), calibrated visibility ($V_c$), and error in $V_c$ ($\sigma_{Vc}$) for each exoplanet host star observed. It shows the information for one star here as an example, and the full table is available on the electronic version of the \emph{Astrophysical Journal}. 

\section{Angular Diameter Determinations}
\begin{figure}[!h]
  \centering \includegraphics[angle=90,width=1.0\textwidth]
  {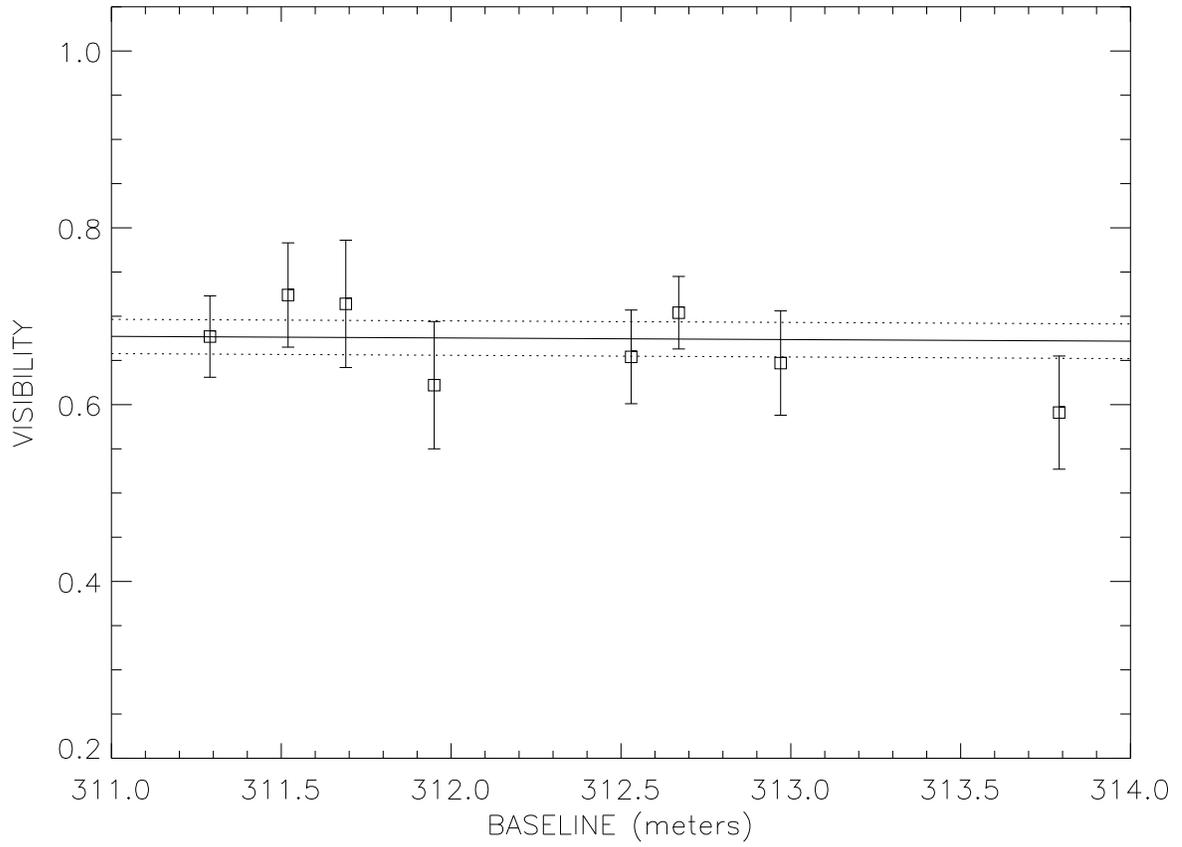}\\
  \caption{Example LD disk diameter fit. HD 3651: Calibrated visibility vs. baseline. The solid line represents the theoretical visibility curve for a star with the best fit $\theta_{\rm LD}$, the dashed lines are the 1$\sigma$ error limits of the diameter fit, the $\Box$s are the calibrated visibilities, and the vertical lines are the measured errors. [See the electronic edition of the \emph{Astrophysical Journal} for Figures 2-24.]}
  \label{HD3651_LDplot}
\end{figure}

The observed quantity of an interferometer is defined as the squared visibility, though we use the unsquared visibility ($V$) in our calculations here. Diameter fits to visibilities were based upon the uniform disk (UD) approximation given by
\begin{equation}
V = \frac{2 J_1(x) }{x},
\end{equation}
where \emph{J}$_1$ is the first-order Bessel function and
\begin{equation}
x = \pi B \theta_{\rm UD} \lambda^{-1},
\end{equation}
where $B$ is the projected baseline at the star's position, $\theta_{\rm UD}$ is the apparent UD angular diameter of the star, and $\lambda$ is the effective wavelength of the observation\footnote{Because the flux distributions for these stars in the $K^\prime$-band are in the Rayleigh-Jeans tail, there are no significant differences in the effective wavelengths for stars of differing spectral types.} \citep{1992ARAandA..30..457S}. The limb-darkened (LD) relationship incorporating the linear limb darkening coefficient $\mu_{\lambda}$ \citep{1974MNRAS.167..475H} is given by:
\begin{equation}
V = \left( {1-\mu_\lambda \over 2} + {\mu_\lambda \over 3} \right)^{-1}
\times
\left[(1-\mu_\lambda) {J_1(\rm x) \over \rm x} + \mu_\lambda {\left( \frac{\pi}{2} \right)^{1/2} \frac{J_{3/2}(\rm x)}{\rm x^{3/2}}} \right] .
\end{equation}
The limb-darkening coefficient was obtained from \citet{1995A&AS..114..247C} after adopting the $T_{\rm eff}$ and log~$g$ values required for each star observed. The resulting angular diameters and other relevant parameters are listed in Table \ref{calculating_diam}. The average difference between the UD and LD diameters are on the order of a few percent, and the final angular diameters are little affected by the choice of $\mu_{\lambda}$. A 20\% change in $\mu_{\lambda}$ produced at most a 0.6\% difference in the angular diameter calculation, so even if the $T_{\rm eff}$ and log~$g$ are not well constrained and therefore $\mu_{\lambda}$ is not precisely known, the effect on the LD diameter ($\theta_{\rm LD}$) will not be significant. Figure~\ref{HD3651_LDplot} shows an example of a diameter fit to calibrated visibilities. Figures 2-24 are available in the electronic edition of the \emph{Astrophysical Journal}.

It was assumed that the visibility curve went to unity at a baseline of 0 m and this carried certain implications. For instance, it was assumed the exoplanet host star was a single star and did not host an unseen stellar companion. A companion check was performed for the stars by studying any possible systematics in the single-star uniform-disk fit errors and by searching for separated fringe packets \citep{2006SPIE.6268E..90F}, and the stars all appeared to be single. Another assumption was that the calibrator star's angular diameter was known and could be used to calibrate the target star's visibilities. If the calibrated visibilities exceeded 1 for a given dataset, that calibrator was discarded and the target star was observed again with a new calibrator.

For each $\theta_{\rm LD}$ fit, the errors were derived via the reduced $\chi^2$ minimization method: the diameter fit with the lowest $\chi^2$ was found and the corresponding diameter was the final $\theta_{\rm LD}$ for the star. The errors were calculated by finding the diameter at $\chi^2 + 1$ on either side of the minimum $\chi^2$ and determining the difference between the $\chi^2$ diameter and $\chi^2 +1$ diameter.

Table \ref{calculating_diam} lists the parameters $T_{\rm eff}$ and log~$g$ from spectroscopic studies that define the limb-darkening coefficient $\mu_{\lambda}$. $\theta_{\rm UD}$ was converted to $\theta_{\rm LD}$ using $\mu_\lambda$, and the combination of $\theta_{\rm LD}$ and the \emph{Hipparcos} parallax \citep{2007hnrr.book.....V} led to a linear radius for the star. Table \ref{calculating_diam} also includes LD diameters estimated from SED fits ($\theta_{\rm SED}$) as a comparison to the measured diameters. The sources for the photometry used are listed in Table~\ref{photometry} and is available online. $R_{\rm standard}$ represents the radius expected from the spectral type listed in the second column.


\subsection{Estimated versus Measured Stellar Diameters}
To check the correspondence between the estimated and measured diameters, Figure \ref{SED_LDdiams} plots $\theta_{\rm LD}$ versus $\theta_{\rm SED}$. At diameters $\gtrsim$0.6~mas, the errors for $\theta_{\rm LD}$ become smaller than those for $\theta_{\rm SED}$. This is to be expected, as the smaller diameters are nearing the resolution limit of the CHARA Array, and the uncertainties will be larger for these measurements.

In order to characterize the scatter in the diameters, the standard deviation $\sigma$ of the quantity ($\theta_{\rm LD} - \theta_{\rm SED}$) was determined and was then divided by $\theta_{\rm SED}$ for each star. Then all the $\sigma$/$\theta_{\rm SED}$ were averaged together, with a resulting value of 12\%. This indicates a fairly good correspondence between the estimated and measured diameters.


\section{Combining Stellar Radii from Interferometry and Eclipsing Binary Systems}
It was of particular interest to combine interferometrically-measured stellar radii and radii determined using other direct means in order to check their compatability. Some of the most precise stellar radii result from measuring detached, double-lined eclipsing binary systems, as described in \citet{1991A&ARv...3...91A}. His sample encompasses all spectral types from O8~V to M1~V and includes one system of two evolved stars. The errors in the radius measurements are $\leq$2\%, and the values are presumed to be valid for single stars.

Figure~\ref{rbv_all} shows the stellar radii measured from eclipsing binaries and the exoplanet host stars' linear radii measured here with errors $<$15\%. The Andersen sample has few G and K-dwarfs and our work helps better populate the low-mass range by tripling the number of stellar radii measurements in the $0.5 \leq (B-V) \leq 1.0$ portion of the plot. The radii measured from eclipsing binaries support the validity of the interferometric measurements. Though 21 stars measured here have linear radii errors $<$15\%, 19 are shown in Figure~\ref{rbv_all}. The remaining targets are HD~59686, a K2~III star, and HD~104985, a G9~III star. The figure demonstrates that many stars in our and the Andersen sample are post-zero-age-main-sequence (ZAMS) objects.


\subsection{Separating Dwarfs and Subgiants} \label{dwarfs_subgiants}
Interferometrically-derived radii may reveal the beginnings of post-main-sequence evolution for stars previously classified as dwarfs. Figure \ref{exsolhr} plots the stars listed in Table~\ref{calculating_diam} on a color-magnitude plot except for the giants in the sample (HD~13189, HD~59686, and HD~104985). Over half the stars lie on a fairly well-defined main-sequence (M-S) and their measured radii generally match the expected values (see Figure~\ref{exsolhr}). Because there will be some spread in the M-S due to stars having a non-zero age, we are concentrating on the more evolved cases below.

Five of the stars were previously classified as subgiants and, as expected, lie well off the M-S. HD~10697, HD~38529, HD~177830, and HD~190228 were labeled as subgiants by the papers listed in Table~\ref{calculating_diam}, and observations confirm the classification. The fifth star, HD~11964, was given a G5 spectral type with no luminosity class. Its measured radius is over twice that expected for a G5 dwarf, and so is most likely a subgiant.

Another group of stars previously classified as dwarfs show measured radii that substantially exceed what is expected from the stars' spectral types as given by \citet{2000asqu.book.....C} and show indications of post M-S evolution. These stars are:

\emph{HD~19994:} \citet{2004AandA...415..391M} classified this star as an F8~V star when the planetary system was discovered. Its measured radius is $\sim$60\% larger than that expected for a standard F8 dwarf star.

\emph{HD~23596:} No luminosity class was assigned to HD~23596 by the \emph{SIMBAD Astronomical Database}, only a spectral type of F8. Its measured radius is $\sim$75\% larger than that of an F8 dwarf.

\emph{HD~117176:} \citet{1996ApJ...464L.147M} labeled this star as G4~V and the associated radius for that spectral classification is $0.92~R_\odot$. However, the measured radius is well over twice that value and it is placed next to HD~10697, a known subgiant, on the color-magnitude diagram.

\emph{HD~190360:} \citet{2003AandA...410.1051N} classified HD~190360 as a G6 IV star and our radius measurement is $\sim$30\% larger than that of a G6 V star. While the star shows no photometric indication of significant evolution off the M-S, the radius measurement is overlarge for a dwarf.

\emph{HD~196885:} This star was labeled as an F8~IV by \citet{2006ApJ...649.1010J} and its measured radius exceeds the expected radius for an F8~V by $\sim$50\%. Though there is no photometric evidence of evolution, its radius is significantly larger than expected if the star was a dwarf.

\section{Conclusion}
We observed 24 exoplanet systems in order to measure the host stars' diameters, obtaining 22 limb-darkened angular diameters with errors $<$15\%. After the LD diameters were converted to linear radii when combined with \emph{Hipparcos} parallax, 19 dwarf stars boasted radius errors of $<$15\%, and these were plotted with the radii from the eclipsing binary sample from \citet{1991A&ARv...3...91A}. These new results tripled the number of stars in the $0.5 \leq (B-V) \leq 1.0$ range with known radii. Three giants, 5 subgiants, 11 dwarfs, and 5 moderately evolved stars were measured, covering a wide range of evolutionary stages.

\acknowledgements

Many thanks to Chris Farrington for his invaluable assistance in obtaining some of the data used here. The CHARA Array is funded by the National Science Foundation through NSF grants AST-0307562 and AST-0606958 and by Georgia State University through the College of Arts and Sciences and the Office of the Vice President for Research. This research has made use of the SIMBAD literature database, operated at CDS, Strasbourg, France, and of NASA's Astrophysics Data System. This publication makes use of data products from the Two Micron All Sky Survey, which is a joint project of the University of Massachusetts and the Infrared Processing and Analysis Center/California Institute of Technology, funded by the National Aeronautics and Space Administration and the National Science Foundation.

\clearpage


\begin{deluxetable}{cccccc}
\tablewidth{0pc}
\tabletypesize{\scriptsize}
\tablecaption{Observing Log.\label{observations}}

\tablehead{
 \colhead{Target} & \colhead{Other} & \colhead{Calibrator} & \colhead{Baseline} & \colhead{Date} & \colhead{Number of} \\
  \colhead{HD} & \colhead{Name} & \colhead{HD} & \colhead{(length)} & \colhead{(UT)} & \colhead{Observations} \\ }
\startdata
3651 & 54 Psc & 4568 & S1-E1 (331 m) & 2005/10/22 & 2 \\
     &        &      &               & 2005/10/24 & 6 \\
\hline
9826 & $\upsilon$ And & 6920 & S1-E1 (331 m) & 2004/01/14 & 13  \\
     &                &      &               & 2004/01/15 & 6  \\
     &                &      & W1-S2 (249 m) & 2007/09/05 & 15  \\
\hline
10697 & 109 Psc & 10477 & S1-E1 (331 m) & 2005/10/23 & 4  \\
      &         &       &               & 2007/09/13 & 2  \\
      &         &       &               & 2007/09/14 & 4  \\
\hline
11964 & $\ldots$ & 13456 & W1-S1 (279 m) & 2005/12/13 & 1  \\
      &   &       &               & 2005/12/16 & 5  \\
\hline
13189 & $\ldots$ & 11007 & S1-E1 (331 m) & 2005/12/12 & 4  \\
      &          &       &               & 2006/08/14 & 4  \\
\hline
19994 & 94 Cet & 19411 & S1-E1 (331 m) & 2005/10/21 & 4  \\
      &        &       &               & 2005/10/27 & 6  \\
      &        &       &               & 2005/12/10 & 6  \\
\hline
20367 & $\ldots$ & 21864 & S1-E1 (331 m) & 2005/12/12 & 5  \\
      &  &       &               & 2007/01/24 & 2  \\
\hline
23596 & $\ldots$ & 22521 & S1-E1 (331 m) & 2007/09/11 & 7  \\
      &   &       &               & 2007/09/14 & 5  \\
\hline
38529 & $\ldots$ & 43318 & S1-E1 (331 m) & 2005/10/22 & 2  \\
      &   &       &               & 2005/10/24 & 2  \\
      &   &       &               & 2005/12/06 & 8  \\
\hline
50554 & $\ldots$ & 49736 & S1-E1 (331 m) & 2005/12/07 & 2  \\
      &   &       &               & 2005/12/12 & 5  \\
\hline
59686 & $\ldots$ & 61630 & S1-E1 (331 m) & 2005/12/06 & 8  \\
      &   &       &               & 2007/04/02 & 9  \\
\hline
75732 & 55 Cnc & 72779 & S1-E1 (331 m) & 2007/03/26 & 5  \\
      &        &       &               & 2007/03/30 & 6  \\
\hline
104985 & $\ldots$ & 97619 & E1-W1 (314 m) & 2007/04/26 & 7 \\
\hline
117176 & 70 Vir & 121107 & S1-E1 (331 m) & 2006/05/20 & 5  \\
       &        &        &               & 2007/04/02 & 6  \\
\hline
120136 & $\tau$ Boo & 121107 & S1-E1 (331 m) & 2007/02/05 & 10 \\
       &            &        &               & 2007/03/25 & 2  \\
       &            &        &               & 2007/03/26 & 5  \\
       &            &        &               & 2007/03/30 & 8  \\
\hline
143761 & $\rho$ CrB & 136849 & S1-E1 (331 m) & 2006/05/19 & 4  \\
       &            &        &               & 2006/06/09 & 1  \\
\hline
145675 & 14 Her & 151044 & S1-E1 (331 m) & 2006/08/11 & 3  \\
       &        &        &               & 2006/08/12 & 7  \\
\hline
177830 & $\ldots$ & 176377 & S1-E1 (331 m) & 2006/06/09 & 1  \\
       &   &        &               & 2006/08/13 & 6  \\
\hline
186427 & 16 Cyg B & 184960 & S1-E1 (331 m) & 2006/08/13 & 6  \\
       &          &        &               & 2007/09/12 & 6  \\
\hline
189733 & $\ldots$  & 190993 & S1-E1 (331 m) & 2006/05/31 & 1 \\
       &   &        &               & 2006/06/01 & 2  \\
       &   &        &               & 2006/06/08 & 1  \\
       &   &        &               & 2006/08/15 & 5  \\
\hline
190228 & $\ldots$ & 190470 & S1-E1 (331 m) & 2006/08/14 & 8  \\
\hline
190360 & $\ldots$ & 189108 & S1-E1 (331 m) & 2006/06/09 & 1  \\
       &   &        &               & 2006/08/11 & 9  \\
\hline
196885 & $\ldots$ & 194012 & S1-E1 (331 m) & 2005/10/27 & 4  \\
       &   &        &               & 2006/08/14 & 5  \\
\hline
217014 & 51 Peg A & 218261 & S1-E1 (331 m) & 2006/08/12 & 7  \\
\enddata
\tablecomments{Observations for HD~189733 were obtained using the $H$-band while all other observations were obtained using the $K'$-band.}
\end{deluxetable}

\clearpage


\begin{deluxetable}{ccccc}
\tablewidth{0pc}
\tabletypesize{\scriptsize}
\tablecaption{Calibrator Stars' Basic Parameters.\label{calibrators}}

\tablehead{
\colhead{  } & 
\colhead{$T_{\rm eff}$} & 
\colhead{ }       & 
\colhead{$\theta_{\rm LD}$}   & 
\colhead{ } \\
 
\colhead{HD} & 
\colhead{(K)} & 
\colhead{log $g$} & 
\colhead{(mas)} & 
\colhead{Ref} \\
}

\startdata

4568 & 6310 & 3.95 & 0.347$\pm$0.006 & 1 \\
6920 & 6026 & 3.67 & 0.543$\pm$0.028 & 1 \\
10477 & 4800 & 2.24 & 0.439$\pm$0.018 & 2 \\
11007 & 6165 & 4.20 & 0.511$\pm$0.025 & 1 \\
13456 & 6760 & 4.00 & 0.380$\pm$0.011 & 1 \\
19411 & 5050 & 2.54 & 0.485$\pm$0.019 & 2 \\
21864 & 4660 & 2.14 & 0.440$\pm$0.018 & 2 \\
22521 & 5783 & 3.96 & 0.377$\pm$0.008  & 3 \\
43318 & 6456 & 4.01 & 0.491$\pm$0.030 & 1 \\
49736 & 6026 & 4.25 & 0.312$\pm$0.006 & 1 \\
61630 & 4400 & 1.94 & 1.116$\pm$0.067 & 2 \\
72779 & 5790 & 2.90 & 0.413$\pm$0.010 & 4 \\
97619 & 4390 & 1.94 & 0.835$\pm$0.083 & 2 \\
121107 & 5450 & 1.74 & 0.686$\pm$0.013 & 2 \\
136849 & 10741 & 4.24 & 0.255$\pm$0.016 & 1 \\
151044 & 6166 & 4.38 & 0.379$\pm$0.012 & 1 \\
176377 & 5888 & 4.47 & 0.358$\pm$0.007 & 1 \\
184960 & 6456 & 4.33 & 0.489$\pm$0.017 & 1 \\
189108 & 4800 & 2.34 & 0.585$\pm$0.051 & 2 \\
190470 & 4968 & 4.50 & 0.340$\pm$0.009 & 5 \\
190993 & 19055 & $\ldots$ & 0.167$\pm$0.035 & 6 \\
194012 & 6309 & 4.36 & 0.412$\pm$0.008 & 1 \\
218261 & 6165 & 4.40 & 0.384$\pm$0.015 & 1 \\

\enddata
\tablecomments{References indicate the sources of the $T_{\rm eff}$ and log~$g$ values. (1) Allende Prieto \& Lambert (1999); (2) \citet{2000asqu.book.....C}. $T_{\rm eff}$ and log~$g$ based on spectral type as listed in the \emph{SIMBAD Astronomical Database}; (3) Soubiran \& Girard (2005); (4) \citet{2001AJ....121.2159G}; (5) \citet{2003AJ....126.2048G}; (6) \citet{2007ApJ...661L.195B}}
\end{deluxetable}

\clearpage


\begin{deluxetable}{cccccc}
\tablewidth{0pc}
\tablecaption{Example of Exoplanet Host Stars' Calibrated Visibilities.\label{exp_calib_visy}}

\tablehead{
 \colhead{Target} & \colhead{ }   & \colhead{$B$} & \colhead{$\Theta$}    & \colhead{ } & \colhead{ } \\
 \colhead{Name}     & \colhead{MJD} & \colhead{(m)}      & \colhead{(deg)} & \colhead{$V_c$} & \colhead{$\sigma V_c$} \\ }
\startdata
HD 3651 & 53665.387 & 311.95 & 174.5 & 0.621 & 0.072 \\
        & 53665.400 & 312.97 & 171.3 & 0.647 & 0.059 \\
        & 53667.328 & 312.67 & 187.9 & 0.703 & 0.041 \\
        & 53667.347 & 311.52 & 183.3 & 0.723 & 0.059 \\
        & 53667.364 & 311.29 & 179.0 & 0.676 & 0.046 \\
        & 53667.377 & 311.69 & 175.7 & 0.713 & 0.072 \\
        & 53667.390 & 312.53 & 172.5 & 0.653 & 0.053 \\
        & 53667.402 & 313.79 & 169.4 & 0.590 & 0.063 \\
\enddata
\tablecomments{The position angle ($\Theta$) is calculated to be east of north. The complete version of this table is in the electronic edition of the \emph{Astrophysical Journal}. A portion is shown here for guidance regarding its content.}
\end{deluxetable}

\clearpage


\begin{deluxetable}{cccccc}
\tablewidth{0pc}
\tablecaption{Exoplanet Host Stars' Calibrated Visibilities.\label{calib_visy}}

\tablehead{
 \colhead{Target} & \colhead{ }   & \colhead{$B$} & \colhead{$\Theta$}    & \colhead{ } & \colhead{ } \\
 \colhead{Name}     & \colhead{MJD} & \colhead{(m)}      & \colhead{(deg)} & \colhead{$V_c$} & \colhead{$\sigma V_c$} \\ }
\startdata
HD 3651 & 53665.387 & 311.95 & 174.5 & 0.621 & 0.072 \\
        & 53665.400 & 312.97 & 171.3 & 0.647 & 0.059 \\
        & 53667.328 & 312.67 & 187.9 & 0.703 & 0.041 \\
        & 53667.347 & 311.52 & 183.3 & 0.723 & 0.059 \\
        & 53667.364 & 311.29 & 179.0 & 0.676 & 0.046 \\
        & 53667.377 & 311.69 & 175.7 & 0.713 & 0.072 \\
        & 53667.390 & 312.53 & 172.5 & 0.653 & 0.053 \\
        & 53667.402 & 313.79 & 169.4 & 0.590 & 0.063 \\

HD 9826 & 53018.136 & 330.55 & 190.3 & 0.366 & 0.052 \\
        & 53018.146 & 330.60 & 188.0 & 0.366 & 0.025 \\
        & 53018.154 & 330.63 & 186.0 & 0.362 & 0.030 \\
        & 53018.165 & 330.64 & 183.4 & 0.403 & 0.027 \\
        & 53018.180 & 330.65 & 269.9 & 0.387 & 0.026 \\
        & 53018.190 & 330.65 & 177.5 & 0.415 & 0.038 \\
        & 53018.200 & 330.64 & 175.1 & 0.373 & 0.038 \\
        & 53018.211 & 330.61 & 172.6 & 0.395 & 0.062 \\
        & 53018.221 & 330.56 & 170.2 & 0.355 & 0.038 \\
        & 53018.233 & 330.45 & 167.4 & 0.364 & 0.040 \\
        & 53018.244 & 330.28 & 164.9 & 0.403 & 0.043 \\
        & 53018.254 & 330.05 & 162.7 & 0.427 & 0.051 \\
        & 53018.263 & 329.73 & 160.6 & 0.346 & 0.038 \\
        & 53019.175 & 330.65 & 180.5 & 0.337 & 0.045 \\
        & 53019.186 & 330.65 & 177.9 & 0.343 & 0.043 \\
        & 53019.196 & 330.64 & 175.5 & 0.387 & 0.038 \\
        & 53019.207 & 330.62 & 172.9 & 0.370 & 0.024 \\
        & 53019.216 & 330.57 & 170.6 & 0.329 & 0.033 \\
        & 53019.255 & 329.93 & 161.8 & 0.347 & 0.039 \\
        & 54348.292 & 242.92 & 177.3 & 0.483 & 0.077 \\
        & 54348.302 & 243.11 & 174.3 & 0.529 & 0.058 \\
        & 54348.311 & 243.43 & 171.4 & 0.537 & 0.050 \\
        & 54348.322 & 243.88 & 168.4 & 0.563 & 0.071 \\
        & 54348.334 & 244.57 & 164.7 & 0.608 & 0.060 \\
        & 54348.344 & 245.24 & 161.7 & 0.649 & 0.060 \\
        & 54348.355 & 245.98 & 158.7 & 0.630 & 0.062 \\
        & 54348.366 & 246.73 & 155.7 & 0.610 & 0.054 \\
        & 54348.376 & 247.46 & 152.8 & 0.558 & 0.044 \\
        & 54348.397 & 248.67 & 147.4 & 0.577 & 0.046 \\
        & 54348.408 & 249.12 & 144.6 & 0.615 & 0.050 \\
        & 54348.419 & 249.36 & 141.9 & 0.618 & 0.054 \\
        & 54348.435 & 249.26 & 138.2 & 0.721 & 0.085 \\
        & 54348.456 & 248.12 & 133.5 & 0.652 & 0.093 \\
        & 54348.467 & 246.94 & 131.0 & 0.579 & 0.074 \\

HD 10697 & 53666.427 & 309.50 & 175.2 & 1.043 & 0.090 \\
         & 53666.443 & 310.79 & 171.2 & 0.928 & 0.068 \\
         & 53666.456 & 312.37 & 167.9 & 1.004 & 0.104 \\
         & 53666.470 & 314.46 & 164.6 & 0.832 & 0.075 \\
         & 54356.765 & 317.17 & 226.9 & 0.883 & 0.103 \\
         & 54356.775 & 320.54 & 227.4 & 1.029 & 0.125 \\
         & 54357.783 & 323.64 & 228.1 & 0.851 & 0.077 \\
         & 54357.792 & 325.84 & 228.8 & 0.723 & 0.068 \\
         & 54357.802 & 327.87 & 229.7 & 0.816 & 0.089 \\
         & 54357.810 & 328.95 & 230.4 & 0.791 & 0.061 \\

HD 11964 & 53717.253 & 269.16 & 127.2 & 0.903 & 0.080 \\
         & 53720.237 & 266.29 & 127.2 & 0.839 & 0.085 \\
         & 53720.252 & 271.34 & 127.3 & 0.834 & 0.089 \\
         & 53720.264 & 274.76 & 127.5 & 0.770 & 0.088 \\
         & 53720.279 & 277.28 & 128.1 & 0.843 & 0.137 \\
         & 53720.295 & 278.45 & 128.9 & 0.864 & 0.156 \\

HD 13189 & 53716.270 & 327.09 & 184.4 & 0.607 & 0.056 \\
         & 53716.285 & 326.91 & 180.9 & 0.531 & 0.081 \\
         & 53716.298 & 326.96 & 177.7 & 0.589 & 0.095 \\
         & 53716.312 & 327.21 & 174.3 & 0.575 & 0.130 \\
         & 53961.441 & 326.60 & 216.4 & 0.622 & 0.051 \\
         & 53961.454 & 328.41 & 214.4 & 0.648 & 0.062 \\
         & 53961.467 & 329.65 & 212.2 & 0.643 & 0.073 \\
         & 53961.481 & 330.38 & 209.8 & 0.607 & 0.040 \\

HD 19994 & 53664.336 & 297.29 & 213.9 & 0.588 & 0.068 \\
         & 53664.352 & 288.97 & 211.5 & 0.736 & 0.073 \\
         & 53664.365 & 281.76 & 209.1 & 0.669 & 0.056 \\
         & 53664.378 & 274.84 & 206.6 & 0.598 & 0.062 \\
         & 53670.286 & 312.83 & 217.7 & 0.750 & 0.080 \\
         & 53670.301 & 306.26 & 216.2 & 0.726 & 0.084 \\
         & 53670.316 & 298.87 & 214.4 & 0.690 & 0.055 \\
         & 53670.333 & 290.10 & 211.9 & 0.764 & 0.076 \\
         & 53670.350 & 281.33 & 209.0 & 0.706 & 0.050 \\
         & 53670.367 & 272.38 & 205.5 & 0.792 & 0.048 \\
         & 53714.182 & 305.88 & 216.1 & 0.712 & 0.083 \\
         & 53714.208 & 292.79 & 212.7 & 0.795 & 0.107 \\
         & 53714.220 & 286.27 & 210.7 & 0.787 & 0.078 \\
         & 53714.239 & 276.58 & 207.2 & 0.713 & 0.069 \\
         & 53714.254 & 268.97 & 204.0 & 0.775 & 0.068 \\
         & 53714.266 & 262.95 & 201.0 & 0.832 & 0.096 \\

HD 20367 & 53716.339 & 325.81 & 179.2 & 0.855 & 0.147 \\
         & 53716.353 & 326.00 & 175.8 & 0.934 & 0.162 \\
         & 53716.367 & 326.40 & 172.6 & 0.879 & 0.118 \\
         & 53716.380 & 326.96 & 169.5 & 0.925 & 0.124 \\
         & 53716.393 & 327.66 & 166.4 & 0.822 & 0.174 \\
         & 54124.220 & 325.80 & 270.0 & 0.905 & 0.138 \\
         & 54124.232 & 325.91 & 176.9 & 0.919 & 0.131 \\

HD 23596 & 54354.937 & 317.98 & 233.5 & 0.856 & 0.073 \\
         & 54354.945 & 319.79 & 234.9 & 0.929 & 0.080 \\
         & 54354.953 & 321.52 & 236.4 & 0.967 & 0.068 \\
         & 54354.960 & 322.96 & 237.9 & 0.915 & 0.074 \\
         & 54354.969 & 324.30 & 239.5 & 0.879 & 0.056 \\
         & 54354.977 & 325.43 & 241.2 & 0.826 & 0.060 \\
         & 54354.985 & 326.44 & 242.9 & 0.818 & 0.057 \\
         & 54357.938 & 320.98 & 234.3 & 1.040 & 0.098 \\
         & 54357.954 & 324.23 & 237.3 & 0.968 & 0.077 \\
         & 54357.960 & 325.22 & 238.5 & 0.964 & 0.108 \\
         & 54357.967 & 326.10 & 239.7 & 0.972 & 0.059 \\
         & 54357.973 & 326.90 & 241.0 & 0.945 & 0.068 \\

HD 38529 & 53665.444 & 299.99 & 212.7 & 0.809 & 0.072 \\
         & 53665.457 & 293.81 & 210.6 & 0.866 & 0.101 \\
         & 53667.451 & 294.20 & 210.8 & 0.879 & 0.239 \\
         & 53667.464 & 287.73 & 208.4 & 0.801 & 0.085 \\
         & 53710.300 & 309.43 & 215.5 & 0.889 & 0.086 \\
         & 53710.314 & 303.46 & 213.7 & 0.823 & 0.103 \\
         & 53710.330 & 295.80 & 211.3 & 0.805 & 0.089 \\
         & 53710.344 & 289.03 & 208.9 & 0.726 & 0.105 \\
         & 53710.357 & 282.66 & 206.4 & 0.808 & 0.081 \\
         & 53710.370 & 276.44 & 203.6 & 0.888 & 0.065 \\
         & 53710.384 & 270.21 & 200.2 & 0.895 & 0.087 \\
         & 53710.397 & 265.22 & 196.9 & 0.902 & 0.083 \\

HD 50554 & 53711.523 & 317.33 & 174.2 & 0.874 & 0.127 \\
         & 53711.537 & 318.32 & 170.7 & 0.783 & 0.090 \\
         & 53716.422 & 321.04 & 195.4 & 1.006 & 0.138 \\
         & 53716.435 & 319.48 & 192.2 & 0.905 & 0.091 \\
         & 53716.449 & 318.20 & 189.0 & 0.984 & 0.083 \\
         & 53716.463 & 317.30 & 185.7 & 1.027 & 0.096 \\
         & 53716.479 & 316.73 & 181.7 & 0.956 & 0.150 \\

HD 59686 & 53710.430 & 316.56 & 203.4 & 0.406 & 0.038 \\
         & 53710.445 & 313.18 & 200.3 & 0.411 & 0.046 \\
         & 53710.460 & 309.92 & 196.8 & 0.410 & 0.055 \\
         & 53710.474 & 307.34 & 193.6 & 0.420 & 0.061 \\
         & 53710.487 & 305.25 & 190.4 & 0.440 & 0.045 \\
         & 53710.501 & 303.69 & 187.0 & 0.435 & 0.032 \\
         & 53710.514 & 302.71 & 183.6 & 0.453 & 0.044 \\
         & 53710.527 & 302.35 & 90.2 & 0.446 & 0.037 \\
         & 54192.202 & 302.46 & 182.1 & 0.458 & 0.038 \\
         & 54192.223 & 302.65 & 176.6 & 0.490 & 0.039 \\
         & 54192.235 & 303.49 & 173.5 & 0.448 & 0.035 \\
         & 54192.246 & 304.67 & 170.7 & 0.465 & 0.051 \\
         & 54192.257 & 306.19 & 168.1 & 0.451 & 0.042 \\
         & 54192.268 & 307.97 & 165.5 & 0.462 & 0.032 \\
         & 54192.279 & 310.10 & 163.0 & 0.447 & 0.030 \\
         & 54192.290 & 312.37 & 160.5 & 0.439 & 0.027 \\
         & 54192.300 & 314.78 & 158.2 & 0.445 & 0.022 \\

HD 75732 & 54185.221 & 325.59 & 195.2 & 0.562 & 0.061 \\
         & 54185.234 & 324.51 & 192.1 & 0.579 & 0.065 \\
         & 54185.304 & 322.77 & 175.4 & 0.568 & 0.054 \\
         & 54185.317 & 323.32 & 172.3 & 0.576 & 0.058 \\
         & 54185.330 & 324.09 & 169.2 & 0.527 & 0.038 \\
         & 54189.207 & 325.83 & 195.9 & 0.730 & 0.079 \\
         & 54189.221 & 324.70 & 192.7 & 0.651 & 0.090 \\
         & 54189.235 & 323.75 & 189.5 & 0.743 & 0.078 \\
         & 54189.249 & 323.02 & 186.2 & 0.660 & 0.078 \\
         & 54189.264 & 322.56 & 182.6 & 0.765 & 0.080 \\
         & 54189.278 & 322.47 & 179.1 & 0.625 & 0.079 \\

HD 104985 & 54216.214 & 304.67 & 266.0 & 0.456 & 0.045 \\
          & 54216.249 & 307.95 & 253.8 & 0.525 & 0.046 \\
          & 54216.281 & 310.19 & 242.6 & 0.539 & 0.053 \\
          & 54216.299 & 311.13 & 236.3 & 0.485 & 0.051 \\
          & 54216.315 & 311.80 & 230.7 & 0.477 & 0.056 \\
          & 54216.326 & 312.13 & 227.2 & 0.489 & 0.040 \\
          & 54216.336 & 312.43 & 223.6 & 0.498 & 0.045 \\

HD 117176 & 53875.270 & 300.08 & 194.0 & 0.546 & 0.061 \\
          & 53875.292 & 296.28 & 188.4 & 0.538 & 0.078 \\
          & 53875.316 & 294.27 & 182.2 & 0.599 & 0.073 \\
          & 53875.335 & 294.34 & 177.3 & 0.568 & 0.058 \\
          & 53875.352 & 295.80 & 172.6 & 0.571 & 0.056 \\
          & 54192.359 & 311.20 & 203.9 & 0.680 & 0.069 \\
          & 54192.370 & 308.00 & 201.5 & 0.572 & 0.079 \\
          & 54192.383 & 304.58 & 198.6 & 0.629 & 0.073 \\
          & 54192.429 & 295.83 & 187.5 & 0.469 & 0.048 \\
          & 54192.440 & 294.74 & 184.6 & 0.486 & 0.051 \\
          & 54192.450 & 294.20 & 181.7 & 0.492 & 0.041 \\

HD 120136 & 54136.370 & 321.01 & 222.5 & 0.629 & 0.055 \\
          & 54136.381 & 324.27 & 221.9 & 0.661 & 0.058 \\
          & 54136.391 & 326.49 & 221.3 & 0.651 & 0.063 \\
          & 54136.400 & 328.20 & 220.6 & 0.633 & 0.052 \\
          & 54136.411 & 329.58 & 219.8 & 0.680 & 0.047 \\
          & 54136.423 & 330.44 & 218.7 & 0.649 & 0.060 \\
          & 54136.434 & 330.67 & 217.6 & 0.649 & 0.044 \\
          & 54136.445 & 330.34 & 216.3 & 0.605 & 0.046 \\
          & 54136.456 & 329.54 & 214.9 & 0.630 & 0.044 \\
          & 54136.466 & 328.37 & 213.5 & 0.631 & 0.051 \\
          & 54184.452 & 305.95 & 190.2 & 0.695 & 0.076 \\
          & 54184.470 & 304.03 & 185.5 & 0.668 & 0.064 \\
          & 54185.359 & 323.92 & 209.3 & 0.745 & 0.078 \\
          & 54185.370 & 321.72 & 207.4 & 0.718 & 0.050 \\
          & 54185.382 & 319.17 & 205.2 & 0.735 & 0.066 \\
          & 54185.393 & 316.66 & 202.9 & 0.628 & 0.048 \\
          & 54185.403 & 314.46 & 200.9 & 0.580 & 0.057 \\
          & 54189.375 & 318.31 & 204.4 & 0.656 & 0.087 \\
          & 54189.391 & 314.63 & 201.0 & 0.683 & 0.088 \\
          & 54189.403 & 312.13 & 198.5 & 0.681 & 0.055 \\
          & 54189.413 & 310.24 & 196.4 & 0.736 & 0.074 \\
          & 54189.422 & 308.31 & 193.9 & 0.679 & 0.083 \\
          & 54189.435 & 306.50 & 191.1 & 0.661 & 0.079 \\
          & 54189.444 & 305.25 & 188.8 & 0.651 & 0.069 \\
          & 54189.455 & 304.17 & 186.0 & 0.624 & 0.073 \\

HD 143761 & 53874.389 & 328.56 & 190.3 & 0.692 & 0.061 \\
          & 53874.409 & 327.99 & 185.8 & 0.718 & 0.079 \\
          & 53874.425 & 327.74 & 181.8 & 0.790 & 0.088 \\
          & 53874.444 & 327.78 & 177.3 & 0.884 & 0.096 \\
          & 53895.224 & 329.09 & 212.5 & 0.658 & 0.056 \\

HD 145675 & 53958.259 & 329.81 & 168.4 & 0.902 & 0.054 \\
          & 53958.275 & 329.38 & 164.9 & 0.878 & 0.045 \\
          & 53958.292 & 328.62 & 161.0 & 0.859 & 0.051 \\
          & 53959.168 & 329.99 & 189.3 & 1.096 & 0.123 \\
          & 53959.184 & 330.18 & 185.5 & 1.000 & 0.089 \\
          & 53959.200 & 330.26 & 181.8 & 0.964 & 0.069 \\
          & 53959.215 & 330.26 & 178.1 & 0.990 & 0.070 \\
          & 53959.231 & 330.18 & 174.5 & 0.940 & 0.078 \\
          & 53959.246 & 330.01 & 170.9 & 0.954 & 0.067 \\
          & 53959.261 & 329.70 & 167.4 & 0.808 & 0.064 \\

HD 177830 & 53895.335 & 330.37 & 215.0 & 0.873 & 0.080 \\
          & 53960.191 & 330.06 & 209.5 & 0.897 & 0.074 \\
          & 53960.208 & 328.87 & 206.5 & 0.886 & 0.069 \\
          & 53960.223 & 327.38 & 203.5 & 0.949 & 0.086 \\
          & 53960.240 & 325.46 & 199.9 & 0.740 & 0.097 \\
          & 53960.259 & 323.36 & 195.8 & 0.705 & 0.097 \\
          & 53960.276 & 321.74 & 192.0 & 0.935 & 0.104 \\

HD 186427 & 53960.308 & 325.22 & 190.2 & 0.838 & 0.129 \\
          & 53960.324 & 325.76 & 186.5 & 0.974 & 0.092 \\
          & 53960.337 & 326.01 & 183.3 & 0.981 & 0.111 \\
          & 53960.350 & 326.10 & 90.2 & 0.854 & 0.113 \\
          & 53960.363 & 326.03 & 177.0 & 0.895 & 0.090 \\
          & 53960.377 & 325.77 & 173.7 & 1.035 & 0.087 \\
          & 54355.661 & 319.02 & 244.1 & 0.810 & 0.074 \\
          & 54355.667 & 319.87 & 245.4 & 0.828 & 0.090 \\
          & 54355.673 & 320.66 & 246.8 & 0.828 & 0.079 \\
          & 54355.685 & 322.09 & 249.5 & 0.844 & 0.113 \\
          & 54355.692 & 322.70 & 251.0 & 0.952 & 0.117 \\
          & 54355.698 & 323.25 & 252.4 & 0.848 & 0.113 \\

HD 189733 & 53886.905 & 330.5 & $\ldots$ & 0.851 & 0.071 \\
          & 53887.936 & 327.9 & $\ldots$ & 0.843 & 0.056 \\
          & 53887.958 & 324.9 & $\ldots$ & 0.857 & 0.054 \\
          & 53894.865 & 330.5 & $\ldots$ & 0.869 & 0.034 \\
          & 53962.742 & 326.5 & $\ldots$ & 0.909 & 0.069 \\
          & 53962.761 & 323.8 & $\ldots$ & 0.863 & 0.049 \\
          & 53962.778 & 321.3 & $\ldots$ & 0.877 & 0.045 \\
          & 53962.793 & 319.0 & $\ldots$ & 0.839 & 0.045 \\
          & 53962.824 & 315.5 & $\ldots$ & 0.829 & 0.061 \\

HD 190228 & 53961.239 & 330.20 & 207.4 & 0.886 & 0.043 \\
          & 53961.253 & 329.42 & 204.8 & 0.882 & 0.061 \\
          & 53961.268 & 328.30 & 201.8 & 0.868 & 0.073 \\
          & 53961.282 & 327.09 & 198.8 & 0.920 & 0.106 \\
          & 53961.299 & 325.69 & 195.2 & 0.814 & 0.082 \\
          & 53961.313 & 324.58 & 191.9 & 0.858 & 0.084 \\
          & 53961.328 & 323.60 & 188.3 & 0.809 & 0.080 \\
          & 53961.346 & 322.86 & 184.1 & 0.933 & 0.108 \\

HD 190360 & 53895.487 & 326.62 & 193.4 & 0.754 & 0.049 \\
          & 53958.341 & 325.15 & 187.2 & 0.728 & 0.034 \\
          & 53958.357 & 324.67 & 183.6 & 0.741 & 0.035 \\
          & 53958.372 & 324.51 & 269.8 & 0.741 & 0.037 \\
          & 53958.388 & 324.72 & 176.0 & 0.786 & 0.056 \\
          & 53958.405 & 325.30 & 172.0 & 0.698 & 0.039 \\
          & 53958.421 & 326.15 & 168.2 & 0.657 & 0.052 \\
          & 53958.436 & 327.17 & 164.7 & 0.670 & 0.068 \\
          & 53958.451 & 328.17 & 161.6 & 0.636 & 0.060 \\
          & 53958.465 & 329.11 & 158.6 & 0.673 & 0.043 \\

HD 196885 & 53670.153 & 289.70 & 188.8 & 0.909 & 0.085 \\
          & 53670.169 & 287.85 & 184.6 & 0.783 & 0.075 \\
          & 53670.183 & 287.18 & 180.7 & 0.848 & 0.071 \\
          & 53670.196 & 287.44 & 177.1 & 0.950 & 0.090 \\
          & 53961.161 & 330.09 & 220.6 & 0.850 & 0.077 \\
          & 53961.175 & 330.66 & 219.7 & 0.898 & 0.077 \\
          & 53961.188 & 330.30 & 218.8 & 0.819 & 0.085 \\
          & 53961.201 & 329.05 & 217.5 & 0.926 & 0.063 \\
          & 53961.214 & 327.15 & 216.2 & 0.774 & 0.065 \\

HD 217014 & 53959.290 & 329.68 & 218.7 & 0.677 & 0.070 \\
          & 53959.303 & 330.50 & 217.2 & 0.696 & 0.059 \\
          & 53959.316 & 330.63 & 215.6 & 0.639 & 0.061 \\
          & 53959.329 & 330.08 & 213.7 & 0.656 & 0.037 \\
          & 53959.343 & 328.87 & 211.6 & 0.699 & 0.049 \\
          & 53959.357 & 327.11 & 209.3 & 0.611 & 0.074 \\
          & 53959.372 & 324.94 & 206.7 & 0.713 & 0.056 \\
\enddata
\end{deluxetable}

\clearpage


\begin{deluxetable}{cccccc}
\tablewidth{0pc}
\tabletypesize{\scriptsize}
\tablecaption{Photometric Sources.\label{photometry}}

\tablehead{ \colhead{HD} & \colhead{$U$} & \colhead{$B$} & \colhead{$V$} & \colhead{$R$} & \colhead{$I$} \\ }
\startdata
3651 & 1 & 1 & 1 & 1 & 1  \\
9826 & 1 & 1 & 1 & 1 & 1  \\
10697 & $\ldots$ & 2 & 2 & 3 & 3  \\
11964 & $\ldots$ & 2 & 2 & 3 & 3  \\
13189 & $\ldots$ & 2 & 2 & 3 & 3  \\
19994 & 4 & 4 & 4 & 3 & 3 \\
20367 & $\ldots$ & 2 & 2 & 3 & 3  \\
23596 & $\ldots$ & 2 & 2 & 3 & 3 \\
38529 & $\ldots$ & 2 & 2 & 3 & 3  \\
50554 & $\ldots$ & 2 & 2 & 3 & 3  \\
59686 & $\ldots$ & 2 & 2 & 3 & 3  \\
75732 & $\ldots$ & 2 & 2 & 3 & 3  \\
104985 & 4 & 4 & 4 & 3 & 3  \\
117176 & 1 & 1 & 1 & 1 & 1  \\
120136 & 1 & 1 & 1 & 1 & 1  \\
143761 & 4 & 4 & 4 & 3 & 3  \\
145675 & $\ldots$ & 2 & 2 & 3 & 3 \\
177830 & 5 & 5 & 5 & 3 & 3 \\
186427 & 4 & 4 & 4 & 3 & 3 \\
189733 & $\ldots$ & 2 & 2 & 3 & 3 \\
190228 & 6 & 6 & 6 & 3 & 3 \\
190360 & 4 & 4 & 4 & 3 & 3 \\
196885 & $\ldots$ & 2 & 2 & 3 & 3 \\
217014 & 1 & 1 & 1 & 1 & 1  \\
\enddata
\tablecomments{All $JHK$ photometry were from \citet{2003tmc..book.....C}. Sources: (1) \citet{1978A&AS...34..477M}; (2) \citet{1997ESASP1200.....P}; (3) \citet{2003AJ....125..984M}; (4) \citet{1966CoLPL...4...99J}; (5) {1997yCat.2168....0M}; (6) \citet{2001yCat.5109....0M}}
\end{deluxetable}

\clearpage


\begin{deluxetable}{ccccccccccccc}
\rotate
\tablewidth{1.2\textwidth}
\tabletypesize{\scriptsize}
\tablecaption{Exoplanet Host Star Angular Diameter Measurements.\label{calculating_diam}}

\tablehead{ \colhead{ } & \colhead{Spectral} & \colhead{$T_{\rm eff}$} & \colhead{ } & \colhead{ } & \colhead{$\pi$} & \colhead{$\theta_{\rm SED}$} & \colhead{$\theta_{\rm UD}$} & \colhead{$\theta_{\rm LD}$} & \colhead{$\sigma_{\rm LD}$} & \colhead{$R_{\rm linear}$} &\colhead{$\sigma_{\rm R}$} &\colhead{$R_{\rm standard}$} \\
\colhead{HD} & \colhead{Type} & \colhead{(K)} & \colhead{log $g$} & \colhead{$\mu_\lambda$} &\colhead{(mas)} & \colhead{(mas)} & 
\colhead{(mas)} & \colhead{(mas)} & \colhead{(\%)} & \colhead{($R_\odot$)} &\colhead{(\%)} &\colhead{($R_\odot$)} \\ }
\startdata
3651 & K0 V & 5173 & 4.37 & 0.28 & 90.43$\pm$0.32 & 0.767$\pm$0.078 & 0.773$\pm$0.026 & 0.790$\pm$0.027 & 3 & 0.947$\pm$0.032 & 3 & 0.85 \\
9826 & F8 V & 6212 & 4.26 & 0.24 & 74.14$\pm$0.19 & 1.095$\pm$0.032 & 1.091$\pm$0.009 & 1.114$\pm$0.009 & 1 & 1.631$\pm$0.014 & 1 & 1.2 \\
10697 & G5 IV & 5641 & 4.05 & 0.28 & 30.69$\pm$0.43 & 0.496$\pm$0.015 & 0.475$\pm$0.046 & 0.485$\pm$0.046 & 9 & 1.72$\pm$0.17 & 10 & 0.92 \\
11964 & G5 & 5248$^a$ & 3.82$^a$ & 0.30 & 30.43$\pm$0.60 & 0.553$\pm$0.013 & 0.597$\pm$0.078 & 0.611$\pm$0.081 & 13 & 2.18$\pm$0.29 & 13 & 0.92 \\
13189 & K2 II & 4050$^b$ & 1.74$^b$ & 0.37 & 1.80$\pm$0.73 & 0.783$\pm$0.043 & 0.811$\pm$0.027 & 0.836$\pm$0.028 & 3 & 50.39$\pm$20.51 & 41 & $\sim$20 \\
19994 & F8 V & 6217 & 4.29 & 0.24 & 44.28$\pm$0.28 & 0.693$\pm$0.025 & 0.774$\pm$0.026 & 0.788$\pm$0.026 & 3 & 1.930$\pm$0.067 & 3 & 1.2 \\
20367 & G0 V & 6138 & 4.53 & 0.25 & 37.48$\pm$0.63 & 0.386$\pm$0.014 & 0.400$\pm$0.107 & 0.408$\pm$0.109 & 27 & 1.18$\pm$0.32 & 27 & 1.1 \\
23596 & F8 & 6108 & 4.25 & 0.25 & 19.84$\pm$0.49 & 0.264$\pm$0.008 & 0.374$\pm$0.043 & 0.381$\pm$0.044 & 12 & 2.09$\pm$0.24 & 12 & 1.2 \\
38529 & G4 IV & 5674 & 3.94 & 0.28 & 25.46$\pm$0.40& 0.570$\pm$0.028 & 0.561$\pm$0.048 & 0.573$\pm$0.049 & 9 & 2.44$\pm$0.22 & 9 & 1.1 \\
50554 & F8 & 6026 & 4.41 & 0.26 & 33.44$\pm$0.59 & 0.326$\pm$0.009 & 0.338$\pm$0.098 & 0.344$\pm$0.100 & 29 & 1.11$\pm$0.33 & 29 & 1.2 \\
59686 & K2 III & 4571$^a$ & 2.40$^a$ & 0.34 & 10.33$\pm$0.28 & 1.287$\pm$0.064 & 1.074$\pm$0.011 & 1.106$\pm$0.011 & 1 & 11.62$\pm$0.34 & 3 & $\sim$20 \\
75732 & G8 V & 5279 & 4.37 & 0.30 & 80.55$\pm$0.70 & 0.666$\pm$0.029 & 0.834$\pm$0.024 & 0.854$\pm$0.024 & 3 & 1.150$\pm$0.035 & 3 & 0.90 \\
104985 & G9 III & 4877$^c$ & 2.85$^c$ & 0.31 & 10.30$\pm$0.25 & 0.955$\pm$0.065 & 1.006$\pm$0.022 & 1.032$\pm$0.023 & 2 & 10.87$\pm$0.36 & 3 & $\sim$14 \\
117176 & G4 V & 5560 & 4.07 & 0.28 & 55.59$\pm$0.24 & 0.951$\pm$0.068 & 0.986$\pm$0.023 & 1.009$\pm$0.024 & 2 & 1.968$\pm$0.047 & 2 & 0.92 \\
120136 & F7 V & 6339 & 4.19 & 0.24 & 64.03$\pm$0.20 & 0.853$\pm$0.037 & 0.771$\pm$0.015 & 0.786$\pm$0.016 & 2 & 1.331$\pm$0.027 & 2 & 1.3 \\
143761 & G0 V & 5853 & 4.41 & 0.26 & 58.02$\pm$0.28 & 0.700$\pm$0.049 & 0.673$\pm$0.043 & 0.686$\pm$0.044 & 6 & 1.284$\pm$0.082 & 6 & 1.1 \\
145675 & K0 V & 5311 & 4.42 & 0.30 & 56.89$\pm$0.35 & 0.498$\pm$0.008 & 0.363$\pm$0.043 & 0.371$\pm$0.044 & 12 & 0.708$\pm$0.085 & 12 & 0.85 \\
177830 & K0 IV & 4804 & 3.57 & 0.33 & 16.94$\pm$0.63 & 0.515$\pm$0.023 & 0.455$\pm$0.057 & 0.467$\pm$0.058 & 12 & 2.99$\pm$0.39 & 13 & 0.85 \\
186427 & G2.5 V & 5772 & 4.40 & 0.27 & 47.13$\pm$0.27 & 0.494$\pm$0.019 & 0.417$\pm$0.055 & 0.426$\pm$0.056 & 13 & 0.98$\pm$0.13 & 13 & 1.0 \\
189733 & K1 V & 5051$^d$ & 4.53$^d$ & 0.36 & 51.40$\pm$0.69 & 0.363$\pm$0.011 & 0.366$\pm$0.024 & 0.377$\pm$0.024 & 6 & 0.788$\pm$0.051 & 7 & 0.80 \\
190228 & G5 IV & 5312 & 3.87 & 0.30 & 16.25$\pm$0.64 & 0.375$\pm$0.032 & 0.443$\pm$0.045 & 0.453$\pm$0.046 & 10 & 3.02$\pm$0.33 & 11 & 0.92 \\
190360 & G6 IV & 5584 & 4.37 & 0.28 & 63.07$\pm$0.34 & 0.658$\pm$0.031 & 0.682$\pm$0.019 & 0.698$\pm$0.019 & 3 & 1.200$\pm$0.033 & 3 & 0.92 \\
196885 & F8 IV & 6310$^a$ & 4.32$^a$ & 0.24 & 29.83$\pm$0.48 & 0.365$\pm$0.016 & 0.485$\pm$0.046 & 0.494$\pm$0.046 & 9 & 1.79$\pm$0.17 & 10 & 1.2 \\
217014 & G2-3 V & 5804 & 4.42 & 0.27 & 64.09$\pm$0.38 & 0.665$\pm$0.047 & 0.733$\pm$0.026 & 0.748$\pm$0.027 & 4 & 1.266$\pm$0.046 & 4 & 1.0 \\
\enddata
\tablecomments{All $T_{\rm eff}$ and log~$g$ are from \citet{2004A&A...415.1153S} unless otherwise noted. \\
$^a$\citet{1999A&A...352..555A}; $^b$\citet{2000asqu.book.....C}; $^c$\citet{2005PASJ...57..109T}; $^d$\citet{2006A&A...458..873S} \\
$\mu_\lambda$ values are from \citet{1995A&AS..114..247C} and $\pi$ values are from \citet{2007hnrr.book.....V}. \\
The spectral classes are from the following sources: HD~3651: \citet{2003ApJ...590.1081F}; HD~9826: \citet{1999ApJ...526..916B}; HD~10697: \citet{2000ApJ...536..902V}; HD~11964: \citet{2006ApJ...646..505B}; HD~19994: \citet{2004AandA...415..391M}; HD~20367: \citet{2006ApJ...646..505B}; HD~23596: \citet{2005ApJS..159..141V}; HD~38526: \citet{2001ApJ...551.1107F}; HD~50554: Perrier et al. (2003); HD~59686: \citet{2000asqu.book.....C}; HD~75732: \citet{2002ApJ...581.1375M}; HD~104985: \citet{2003ApJ...597L.157S}; HD~117176: \citet{1996ApJ...464L.147M}; HD~120136: \citet{1997ApJ...474L.115B}; HD~143761: \citet{1997ApJ...483L.111N}; HD~145675: \citet{2003ApJ...582..455B}; HD~177830: \citet{2000ApJ...536..902V}; HD~186427: \citet{1997ApJ...483..457C}; HD~189733: Bouchy et al. (2005); HD~190228: Perrier et al. (2003); HD~190360: \citet{2003AandA...410.1051N}; HD~196885: \citet{2006ApJ...649.1010J}; and HD~217014: \citet{1997ApJ...481..926M}.}
\end{deluxetable}

\clearpage


\clearpage

\begin{figure}[!h]
  \centering \includegraphics[angle=90,width=1.0\textwidth]
  {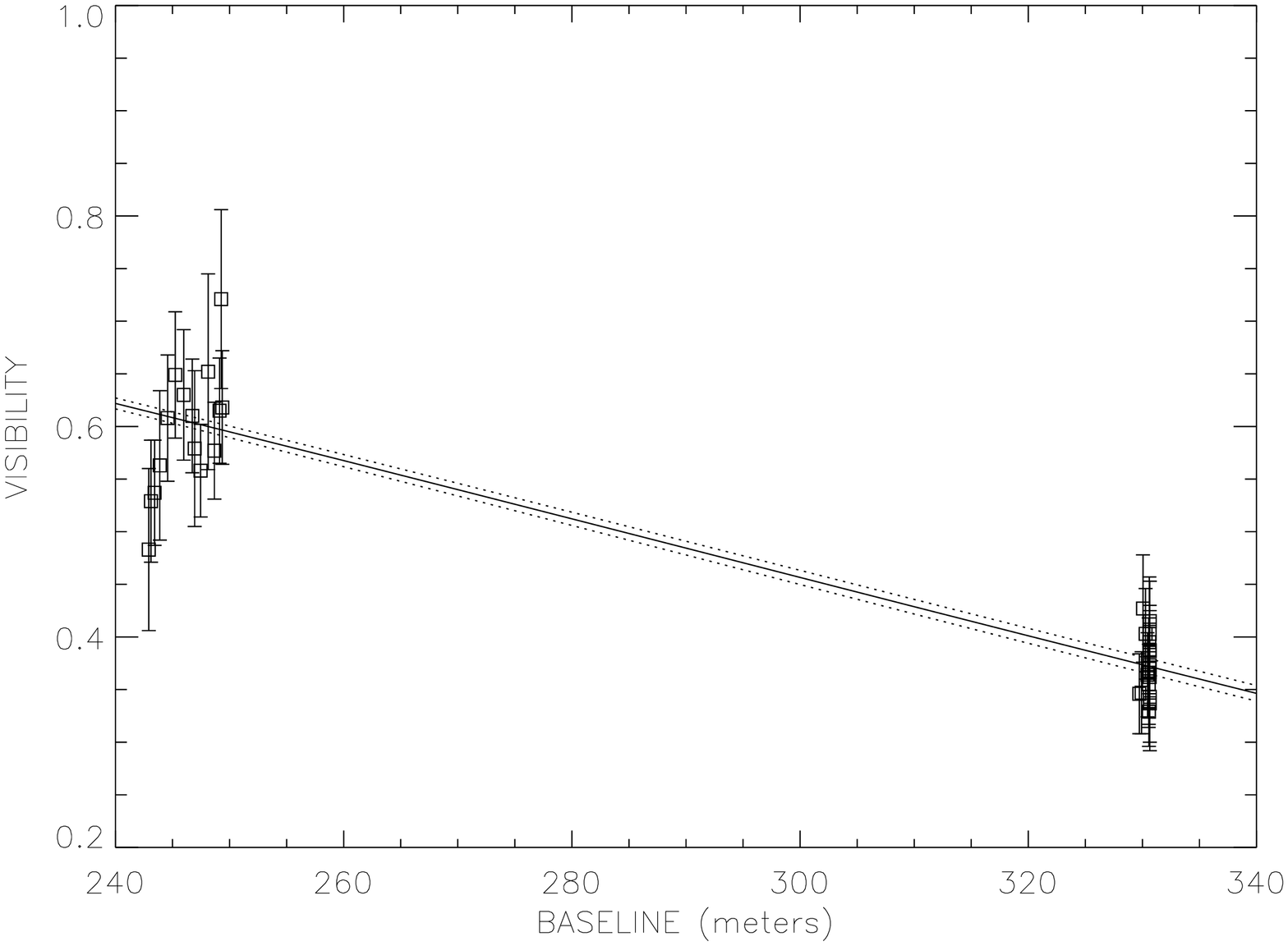}\\
  \caption{HD 9826 limb-darkened disk diameter fit.}
\end{figure}
\clearpage

\begin{figure}[!h]
  \centering \includegraphics[angle=90,width=1.0\textwidth]
  {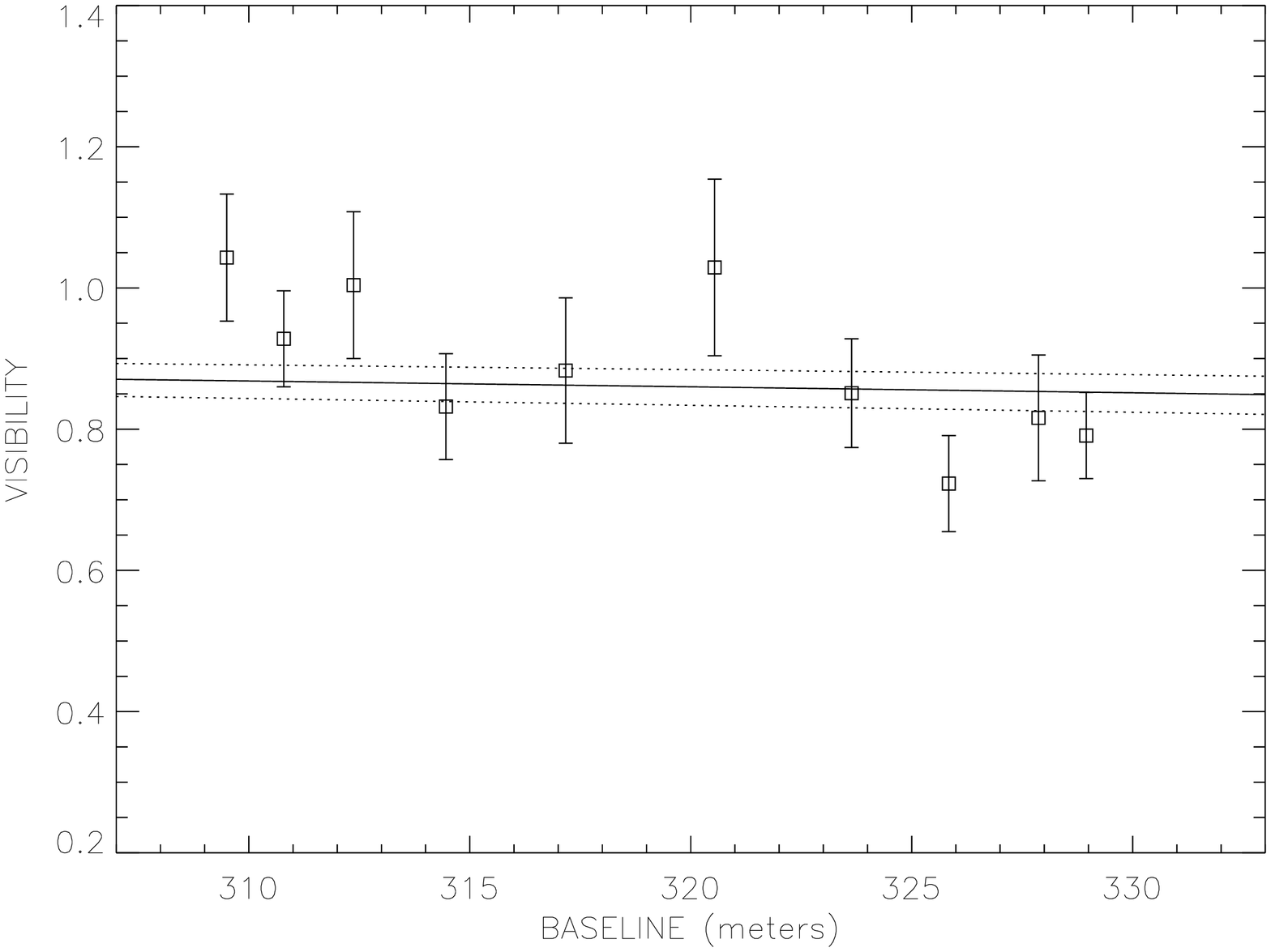}\\
  \caption{HD 10697 limb-darkened disk diameter fit.}
\end{figure}
\clearpage

\begin{figure}[!h]
  \centering \includegraphics[angle=90,width=1.0\textwidth]
  {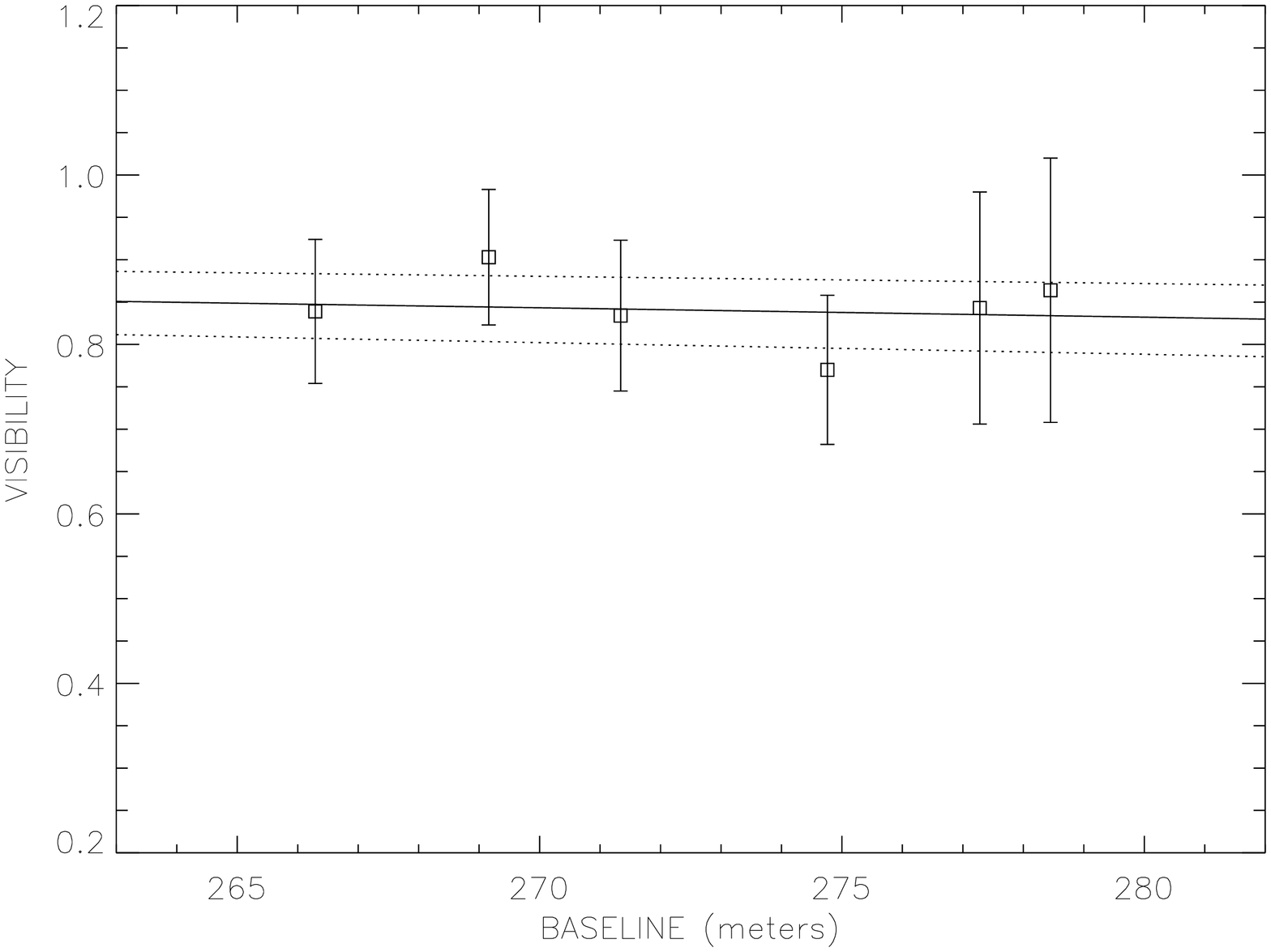}\\
  \caption{HD 11964 limb-darkened disk diameter fit.}
\end{figure}
\clearpage

\begin{figure}[!h]
  \centering \includegraphics[angle=90,width=1.0\textwidth]
  {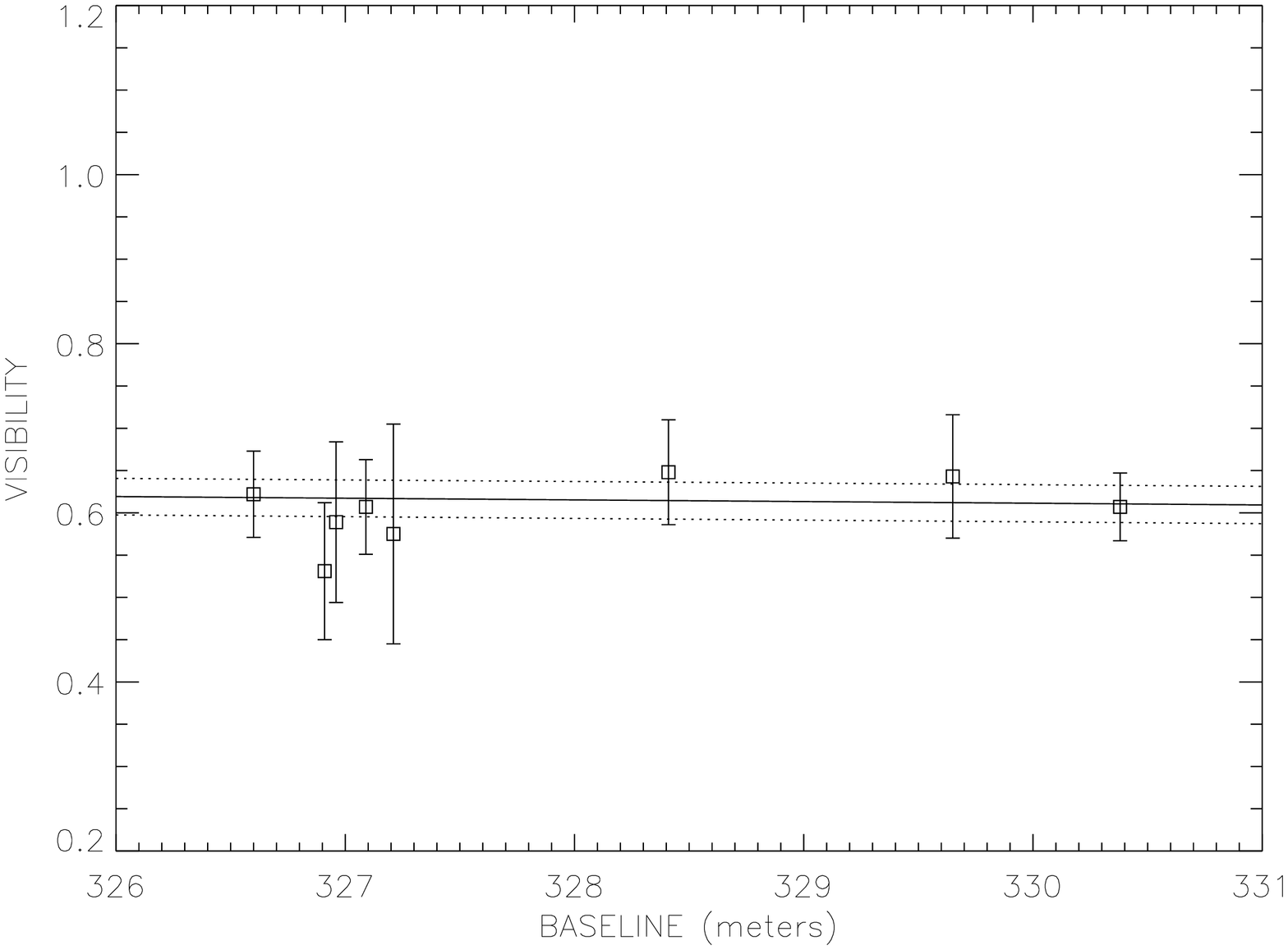}\\
  \caption{HD 13189 limb-darkened disk diameter fit.}
\end{figure}
\clearpage

\begin{figure}[!h]
  \centering \includegraphics[angle=90,width=1.0\textwidth]
  {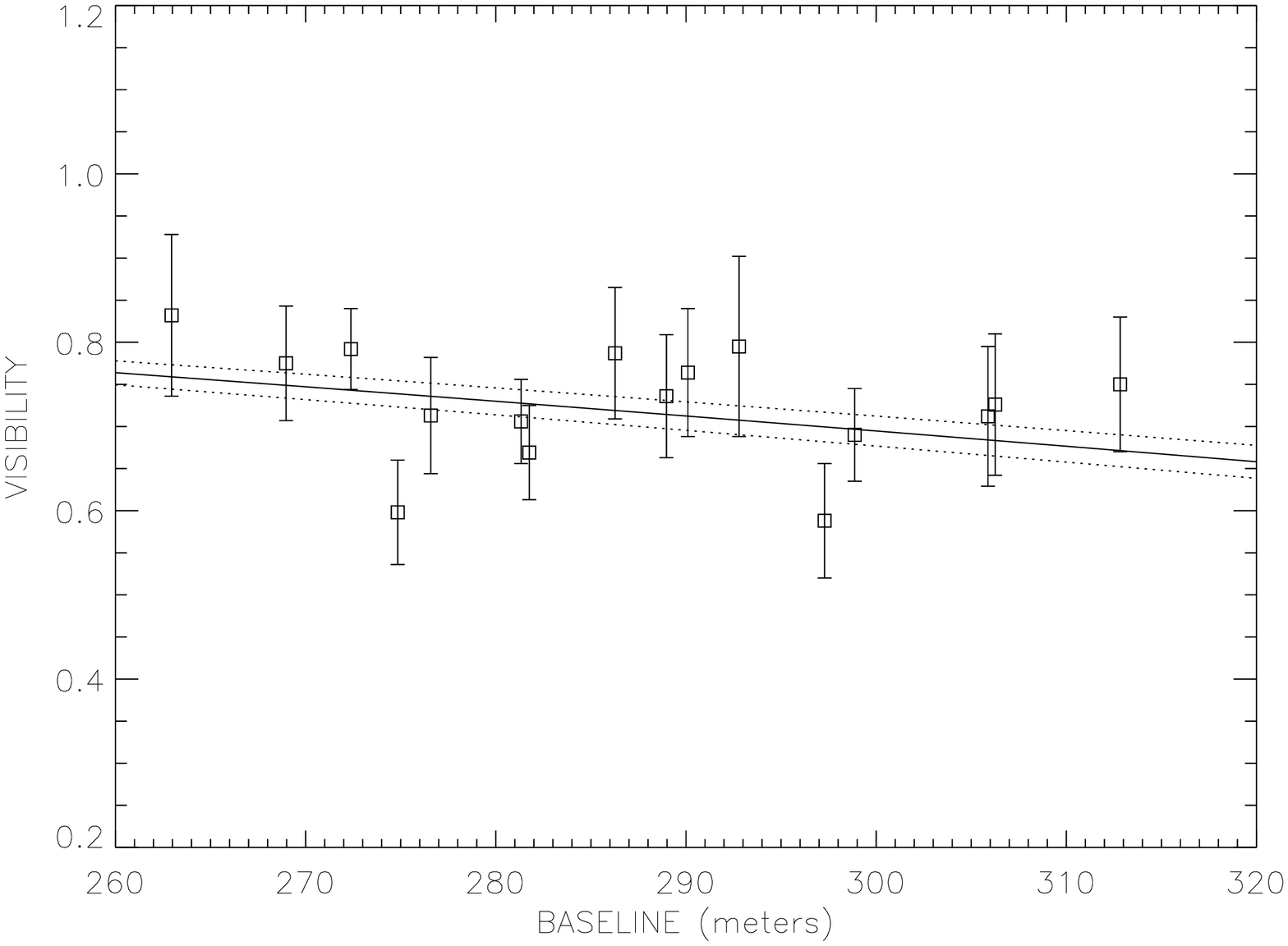}\\
  \caption{HD 19994 limb-darkened disk diameter fit.}
\end{figure}
\clearpage

\begin{figure}[!h]
  \centering \includegraphics[angle=90,width=1.0\textwidth]
  {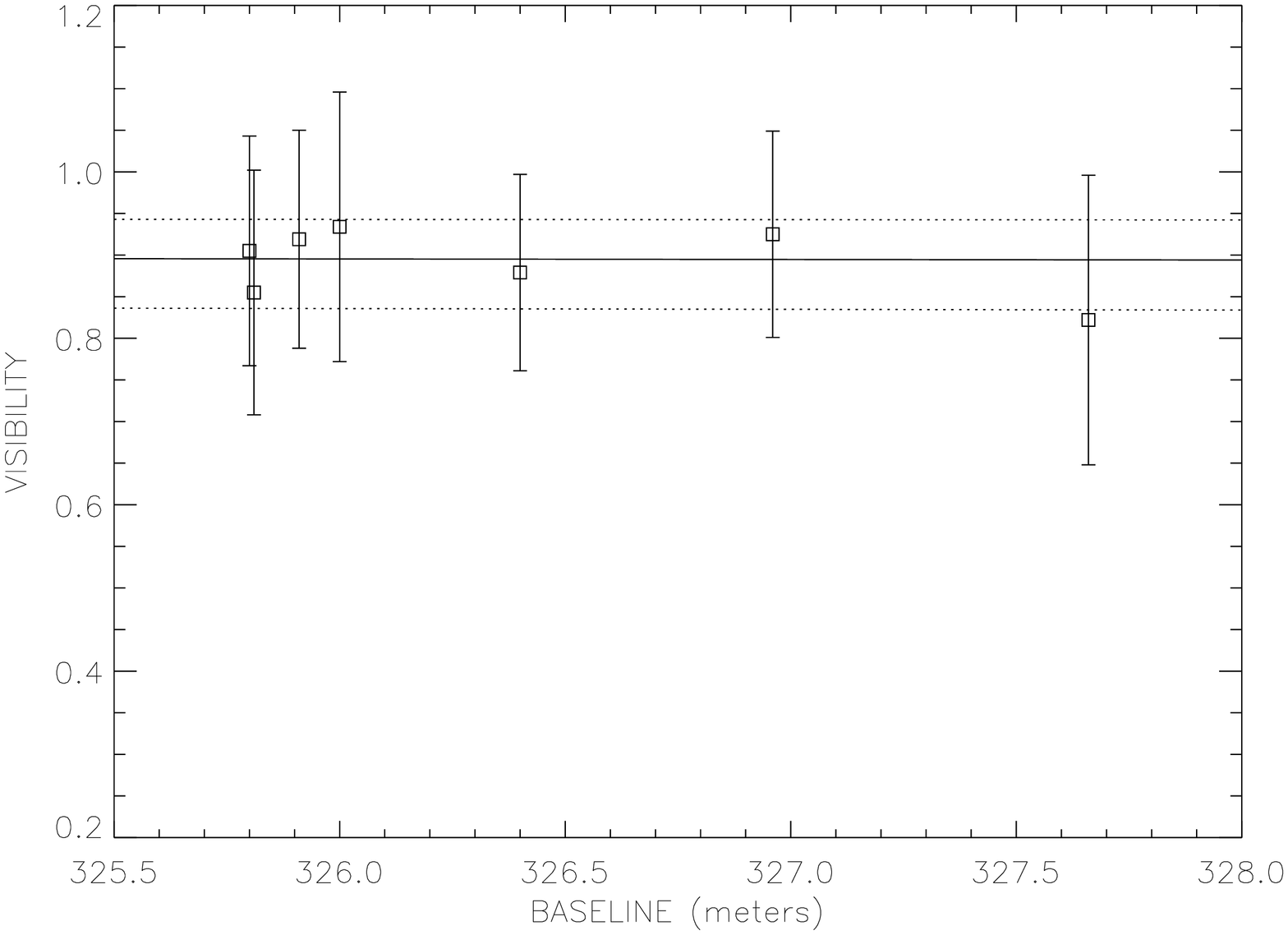}\\
  \caption{HD 20367 limb-darkened disk diameter fit.}
\end{figure}
\clearpage

\begin{figure}[!h]
  \centering \includegraphics[angle=90,width=1.0\textwidth]
  {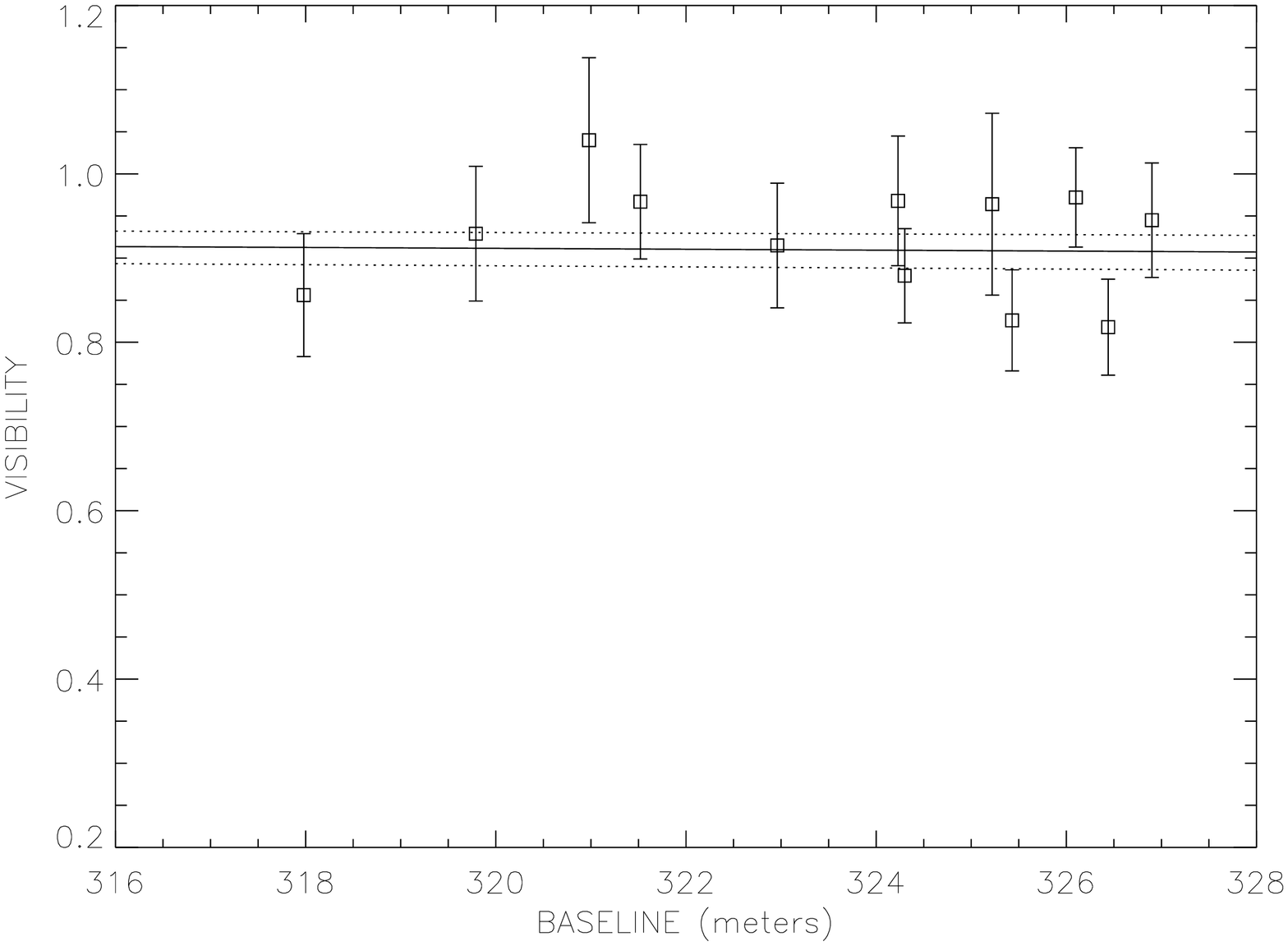}\\
  \caption{HD 23596 limb-darkened disk diameter fit.}
\end{figure}
\clearpage

\begin{figure}[!h]
  \centering \includegraphics[angle=90,width=1.0\textwidth]
  {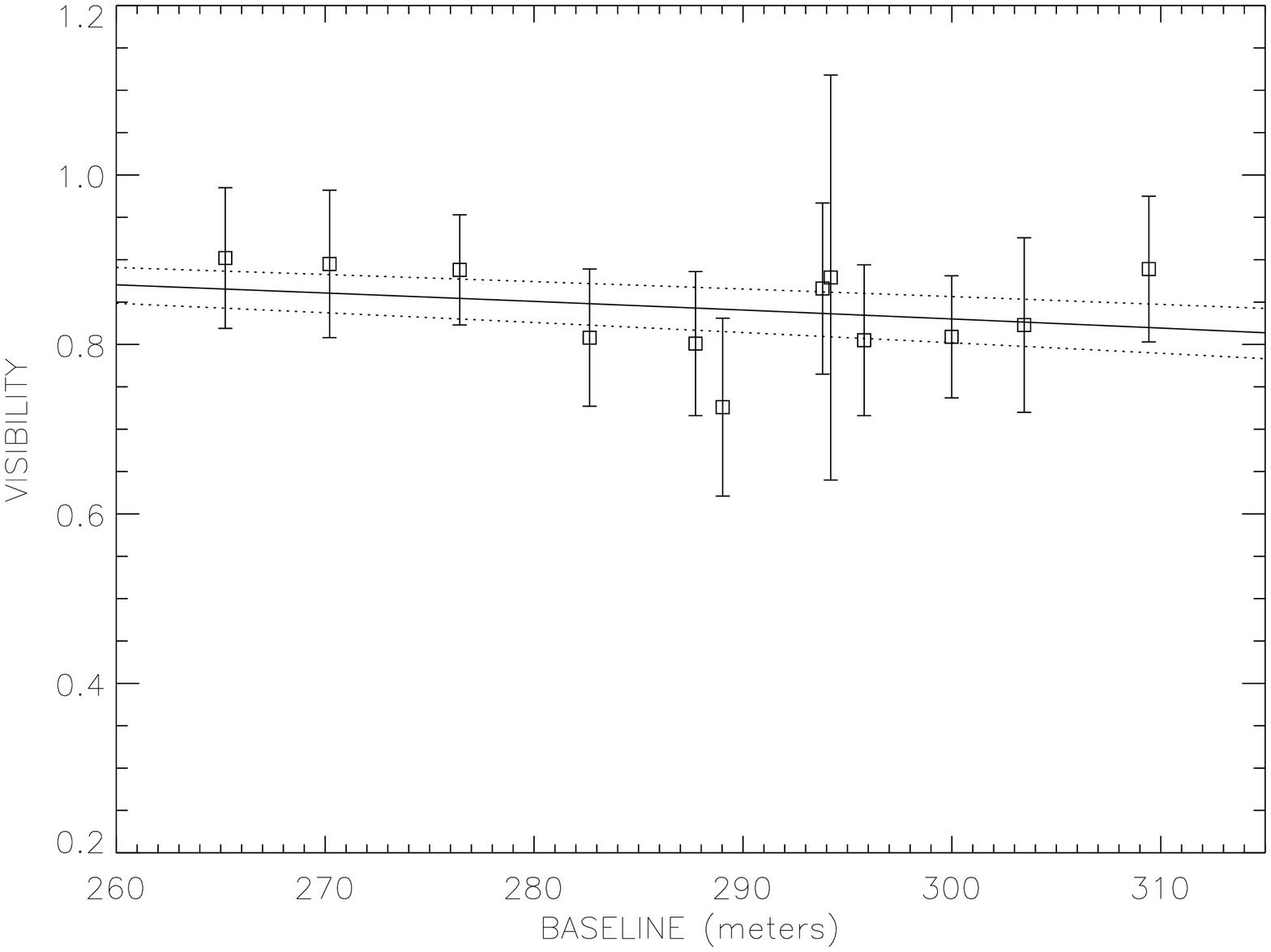}\\
  \caption{HD 38529 limb-darkened disk diameter fit.}
\end{figure}
\clearpage

\begin{figure}[!h]
  \centering \includegraphics[angle=90,width=1.0\textwidth]
  {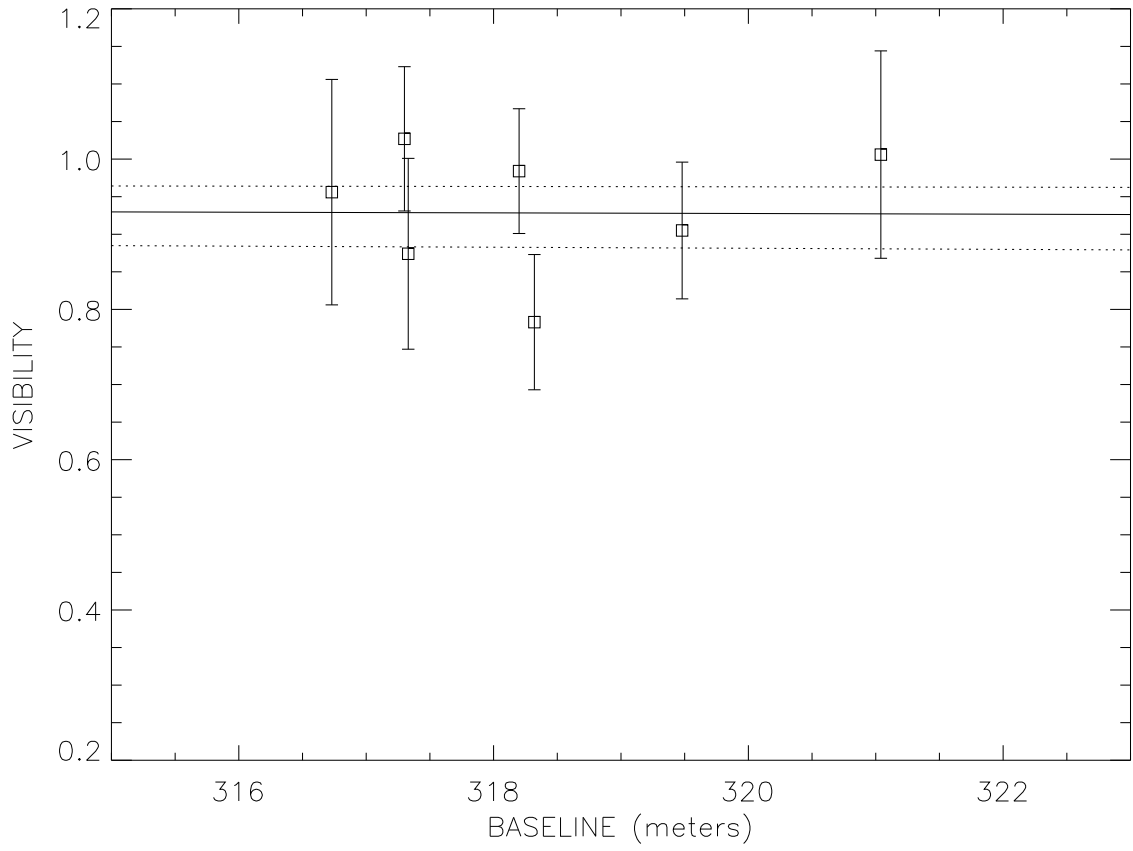}\\
  \caption{HD 50554 limb-darkened disk diameter fit.}
\end{figure}
\clearpage

\begin{figure}[!h]
  \centering \includegraphics[angle=90,width=1.0\textwidth]
  {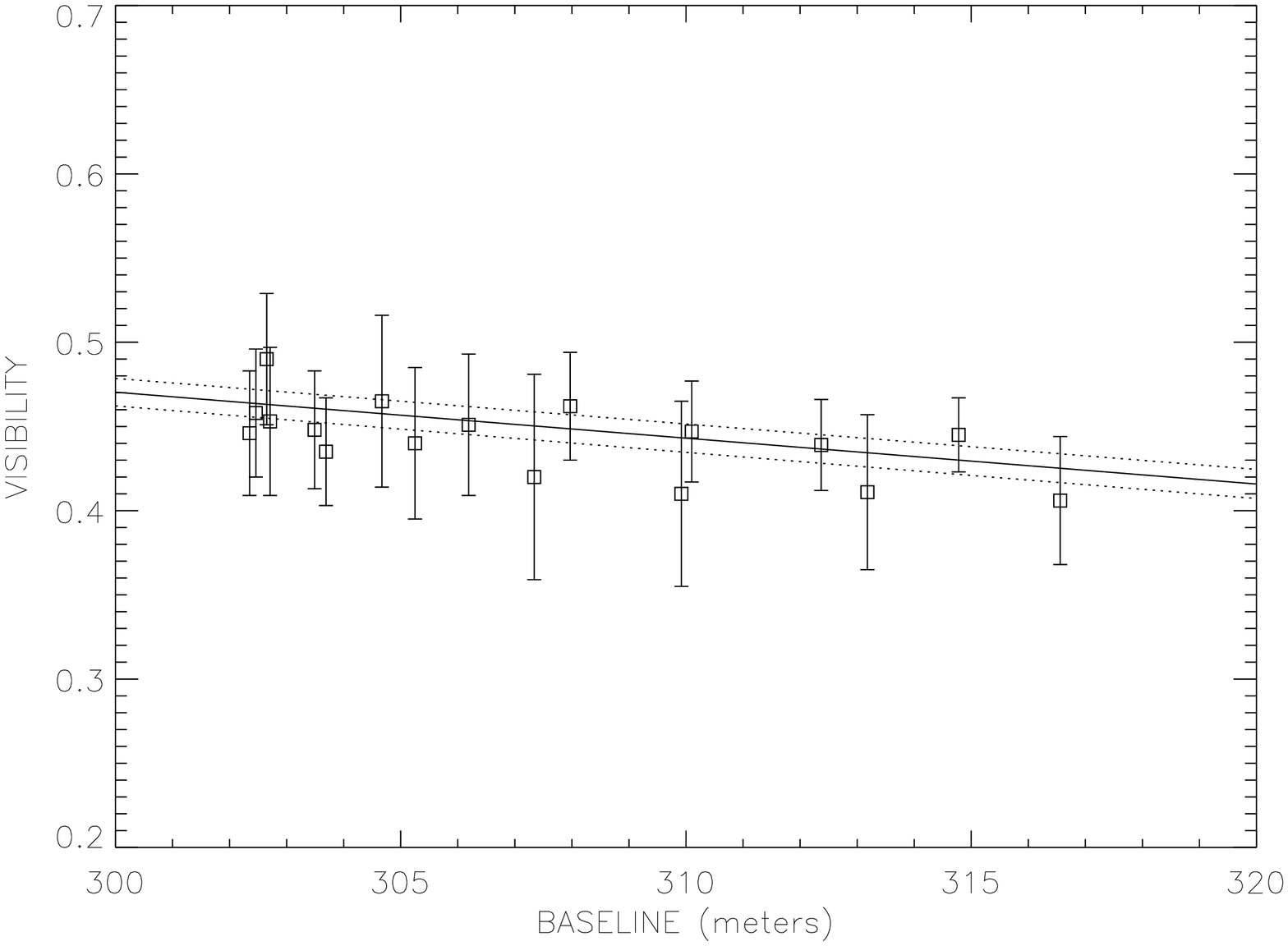}\\
  \caption{HD 59686 limb-darkened disk diameter fit.}
\end{figure}
\clearpage

\begin{figure}[!h]
  \centering \includegraphics[angle=90,width=1.0\textwidth]
  {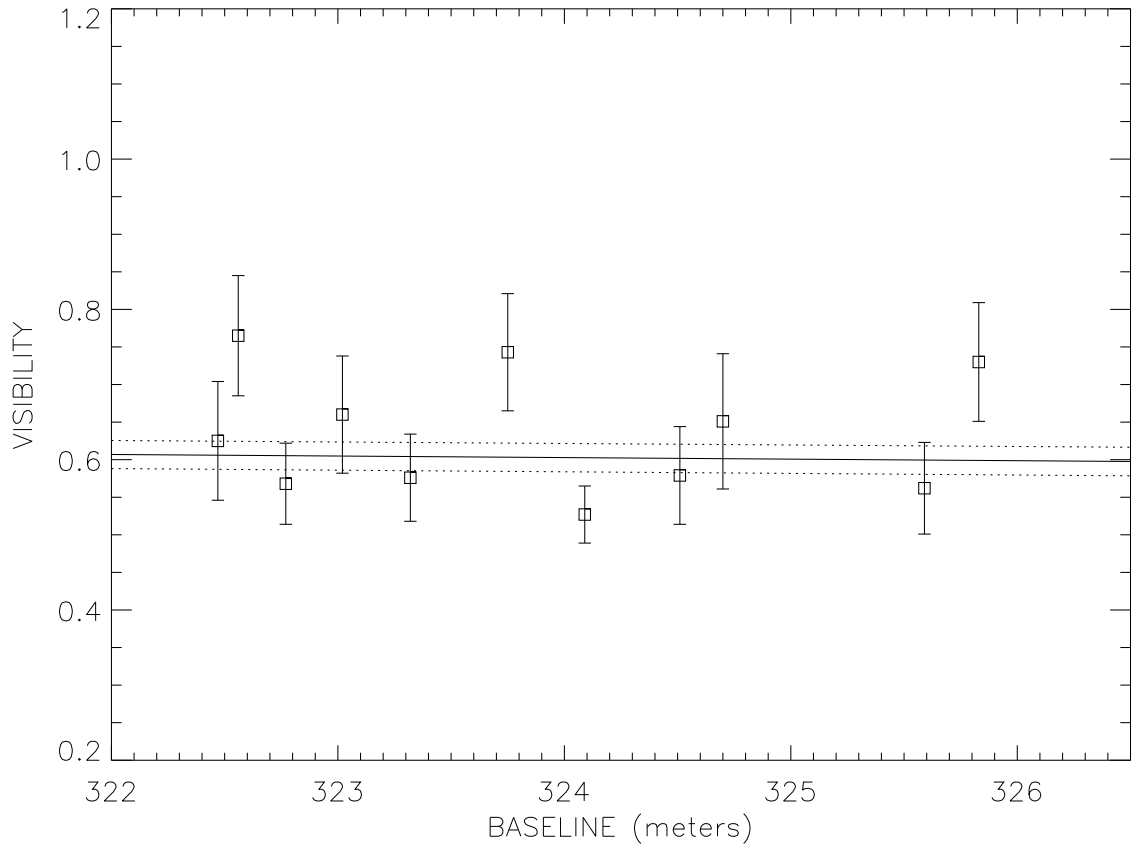}\\
  \caption{HD 75732 limb-darkened disk diameter fit.}
\end{figure}
\clearpage

\begin{figure}[!h]
  \centering \includegraphics[angle=90,width=1.0\textwidth]
  {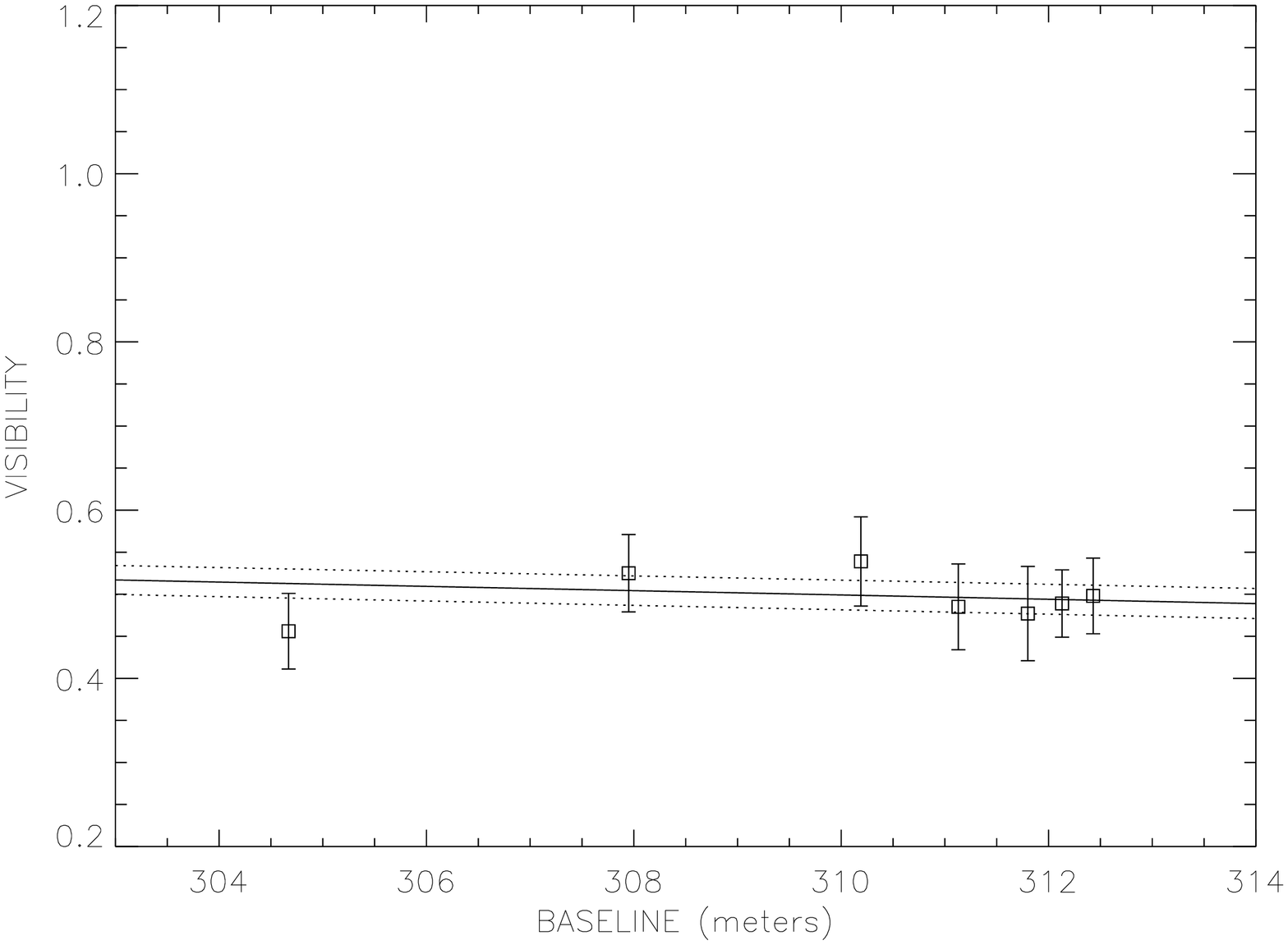}\\
  \caption{HD 104985 limb-darkened disk diameter fit.}
\end{figure}
\clearpage

\begin{figure}[!h]
  \centering \includegraphics[angle=90,width=1.0\textwidth]
  {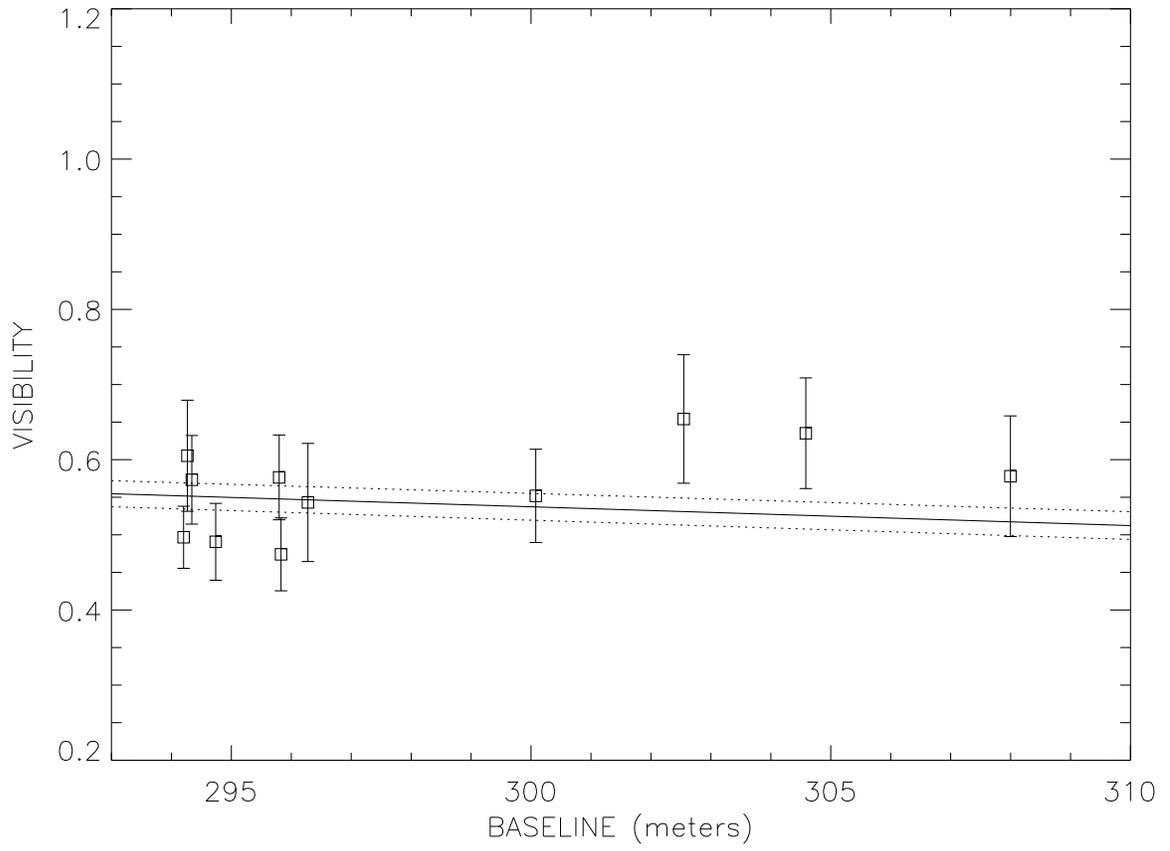}\\
  \caption{HD 117176 limb-darkened disk diameter fit.}
\end{figure}
\clearpage

\begin{figure}[!h]
  \centering \includegraphics[angle=90,width=1.0\textwidth]
  {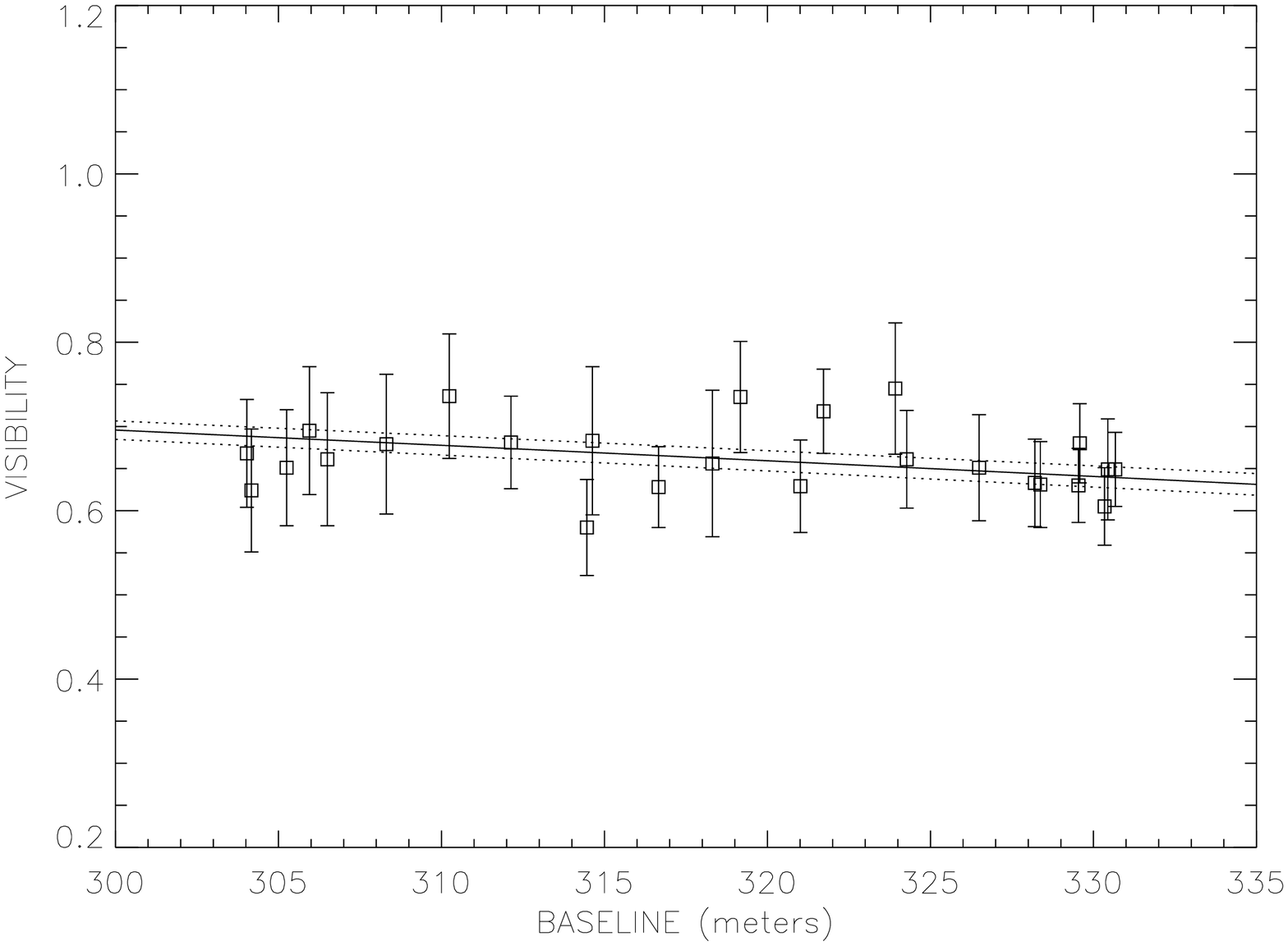}\\
  \caption{HD 120136 limb-darkened disk diameter fit.}
\end{figure}
\clearpage

\begin{figure}[!h]
  \centering \includegraphics[angle=90,width=1.0\textwidth]
  {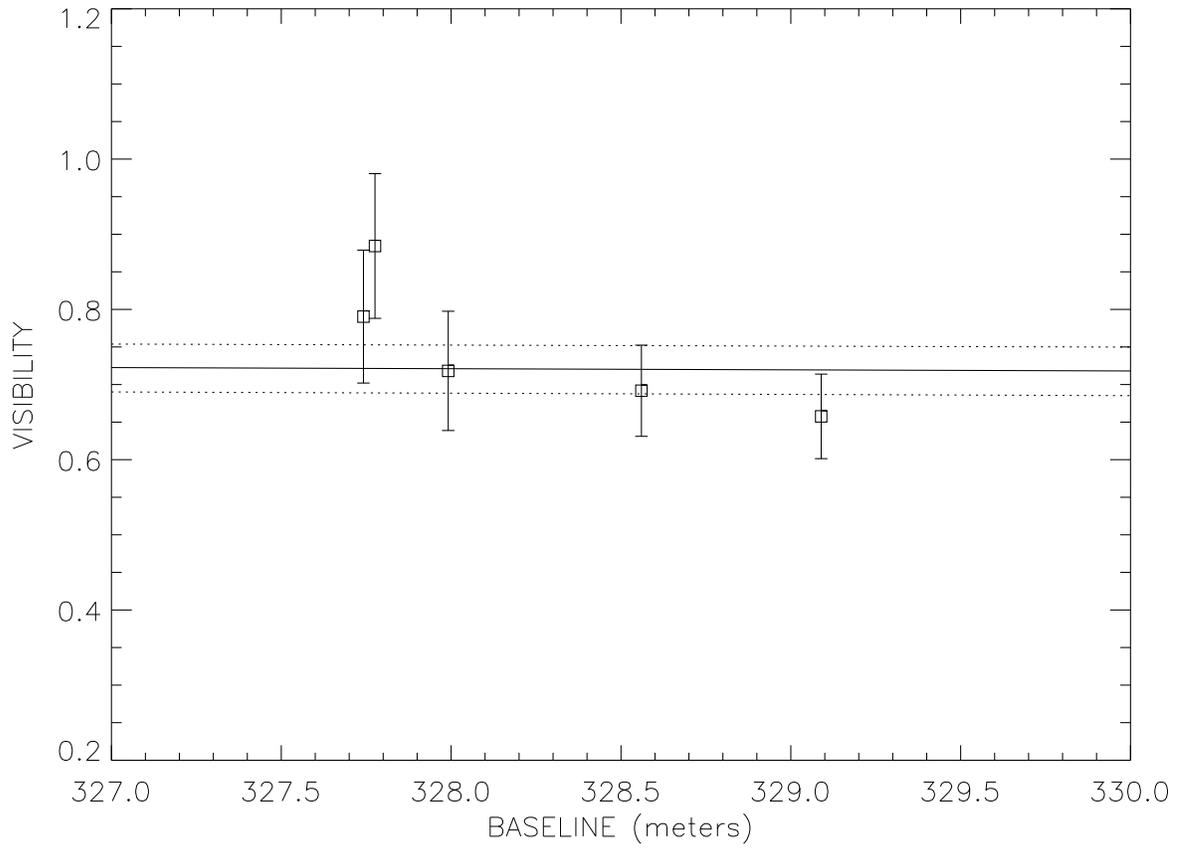}\\
  \caption{HD 143761 limb-darkened disk diameter fit.}
\end{figure}
\clearpage

\begin{figure}[!h]
  \centering \includegraphics[angle=90,width=1.0\textwidth]
  {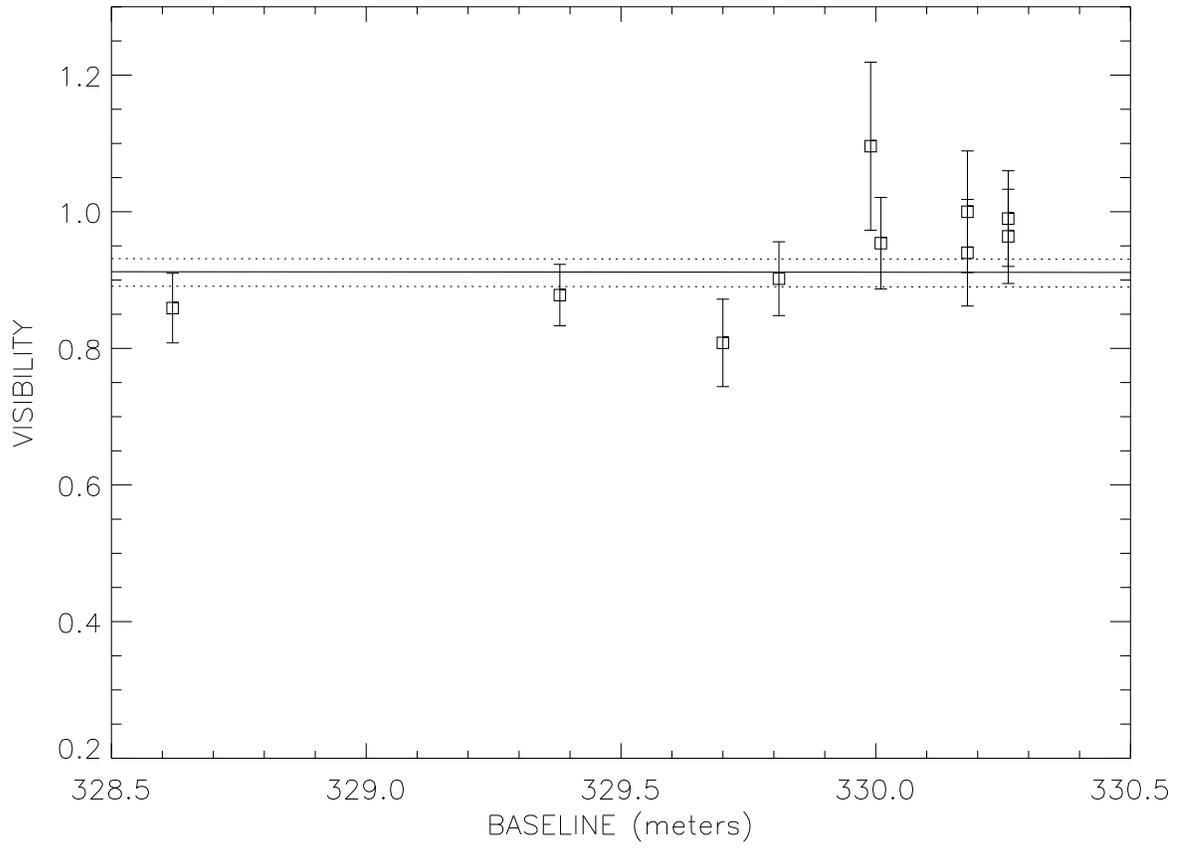}\\
  \caption{HD 145675 limb-darkened disk diameter fit.}
\end{figure}
\clearpage

\begin{figure}[!h]
  \centering \includegraphics[angle=90,width=1.0\textwidth]
  {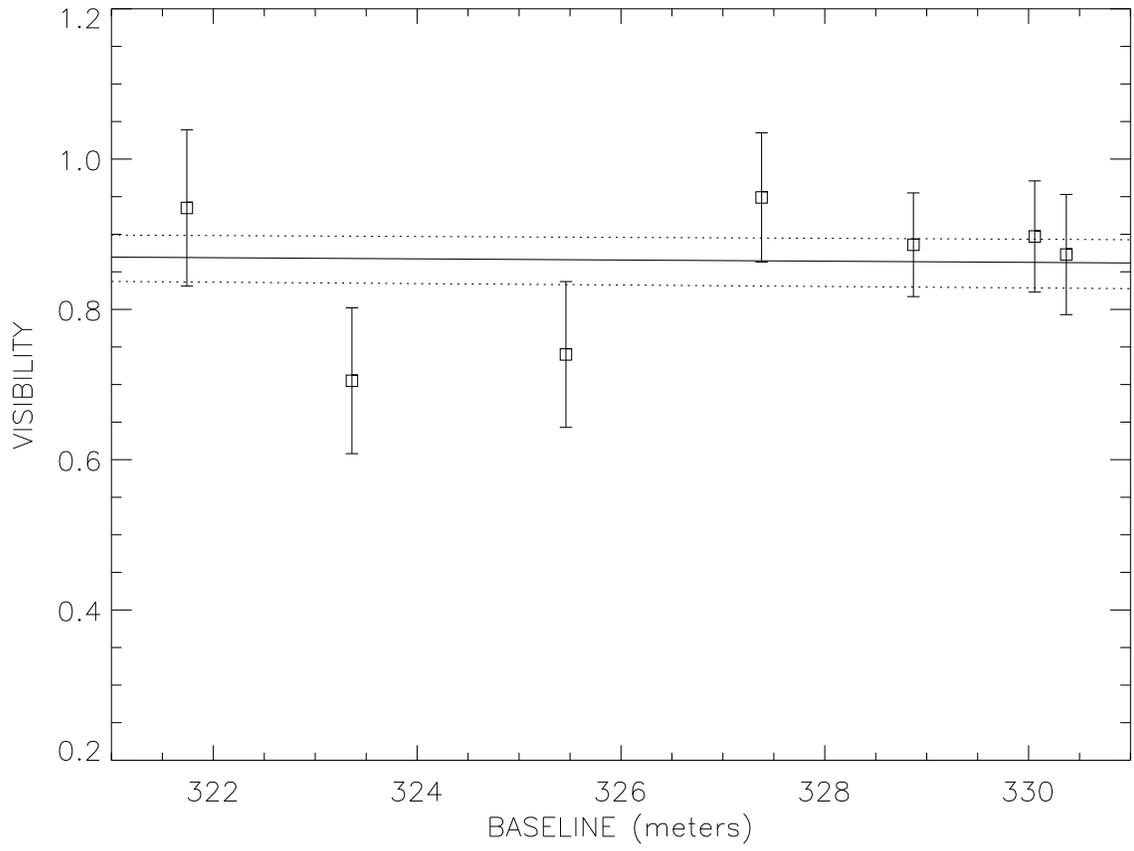}\\
  \caption{HD 177830 limb-darkened disk diameter fit.}
\end{figure}
\clearpage

\begin{figure}[!h]
  \centering \includegraphics[angle=90,width=1.0\textwidth]
  {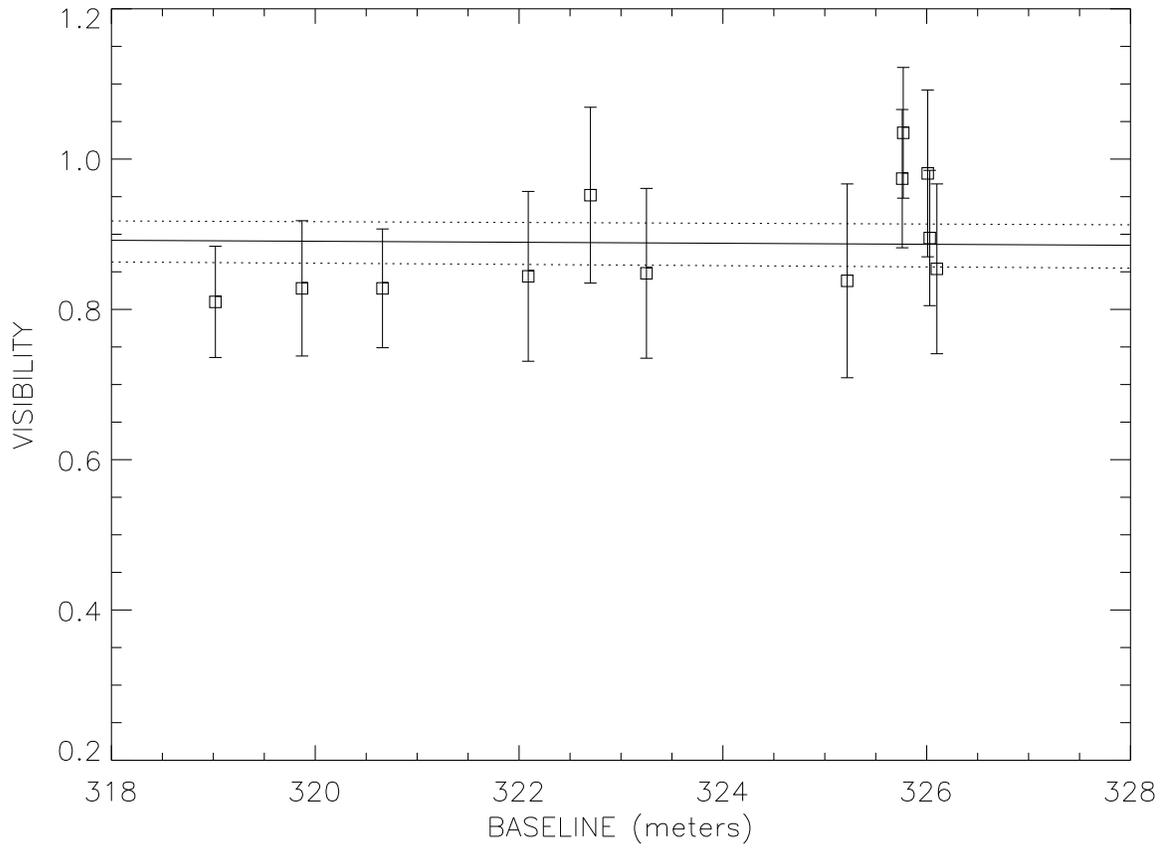}\\
  \caption{HD 186427 limb-darkened disk diameter fit.}
\end{figure}
\clearpage

\begin{figure}[!h]
  \centering \includegraphics[angle=90,width=1.0\textwidth]
  {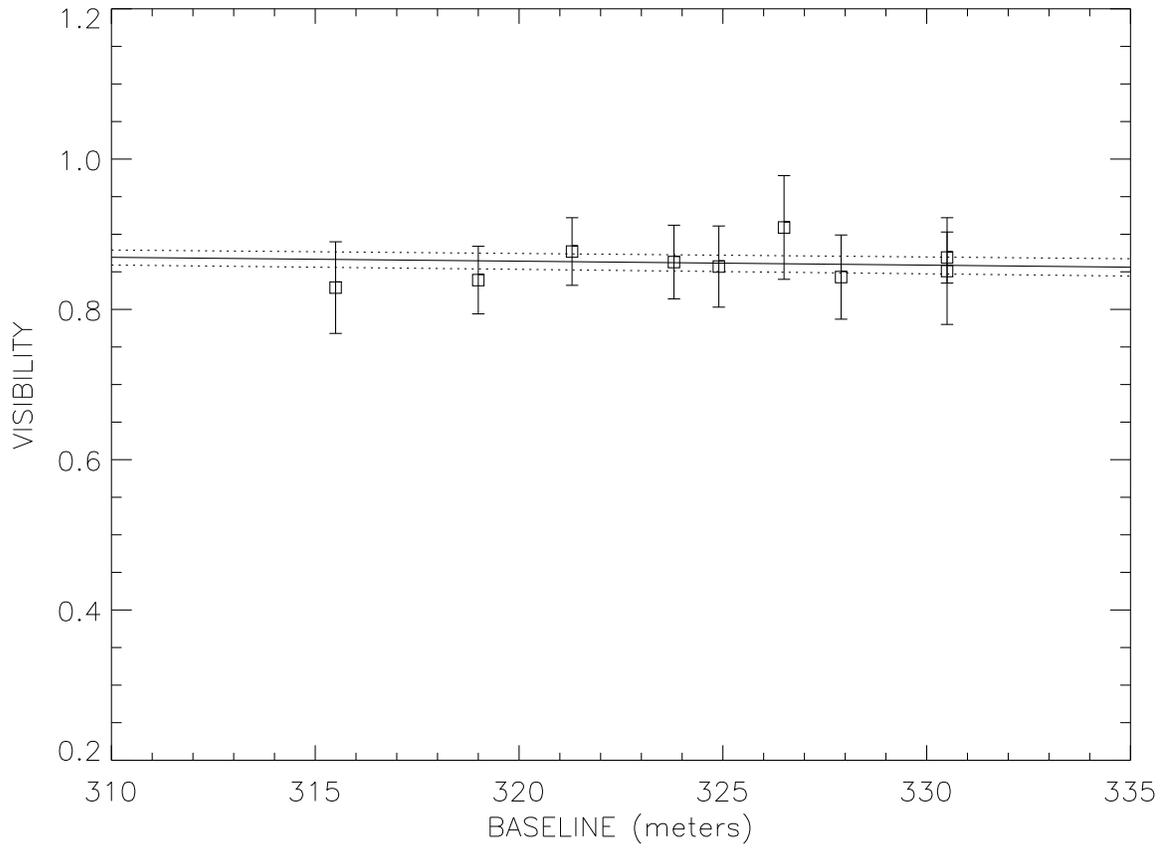}\\
  \caption{HD 189733 limb-darkened disk diameter fit.}
\end{figure}
\clearpage

\begin{figure}[!h]
  \centering \includegraphics[angle=90,width=1.0\textwidth]
  {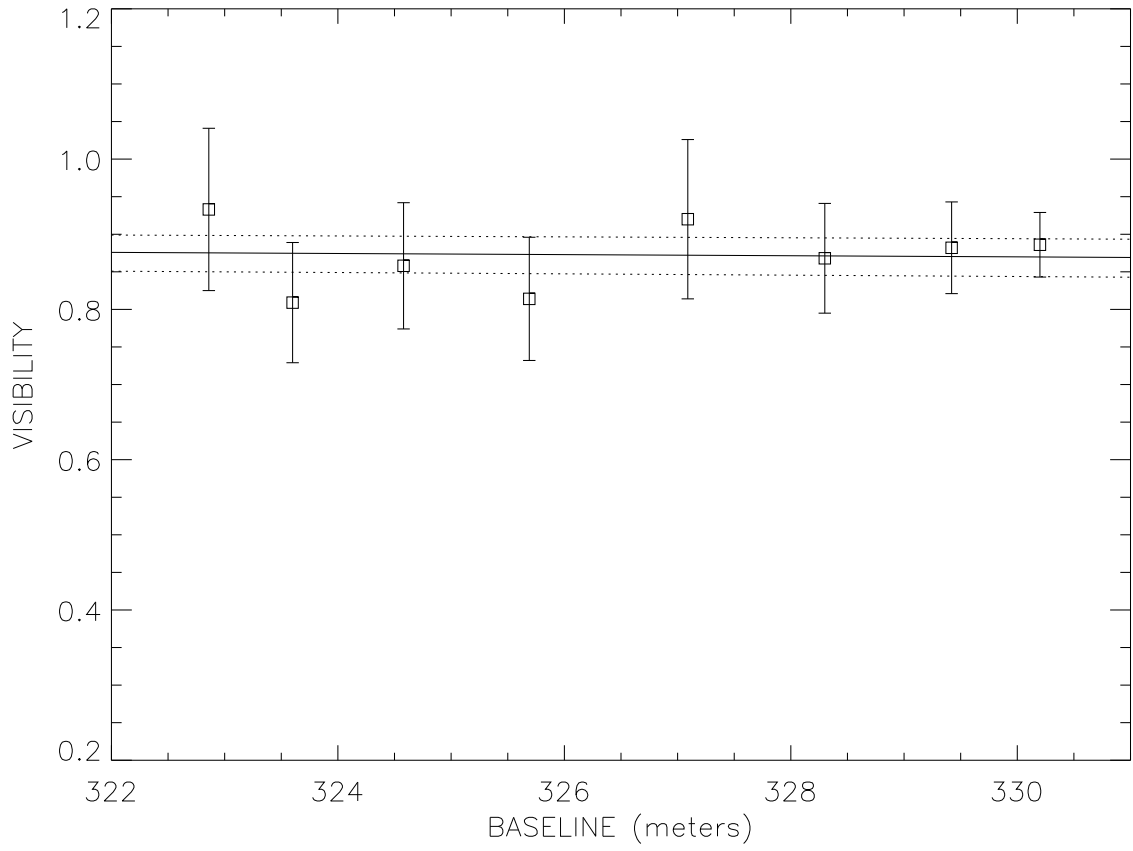}\\
  \caption{HD 190228 limb-darkened disk diameter fit.}
\end{figure}
\clearpage

\begin{figure}[!h]
  \centering \includegraphics[angle=90,width=1.0\textwidth]
  {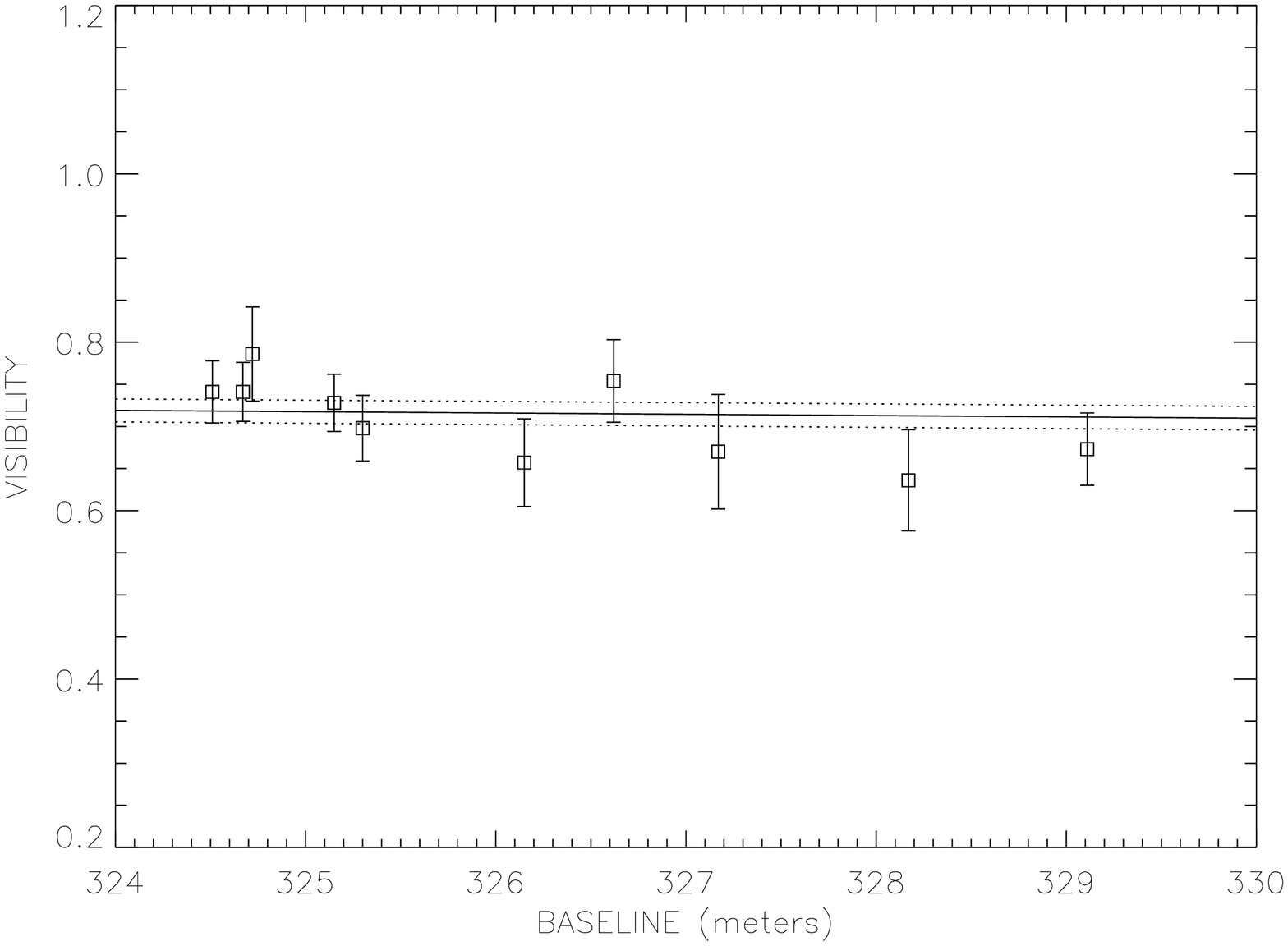}\\
  \caption{HD 190360 limb-darkened disk diameter fit.}
\end{figure}
\clearpage

\begin{figure}[!h]
  \centering \includegraphics[angle=90,width=1.0\textwidth]
  {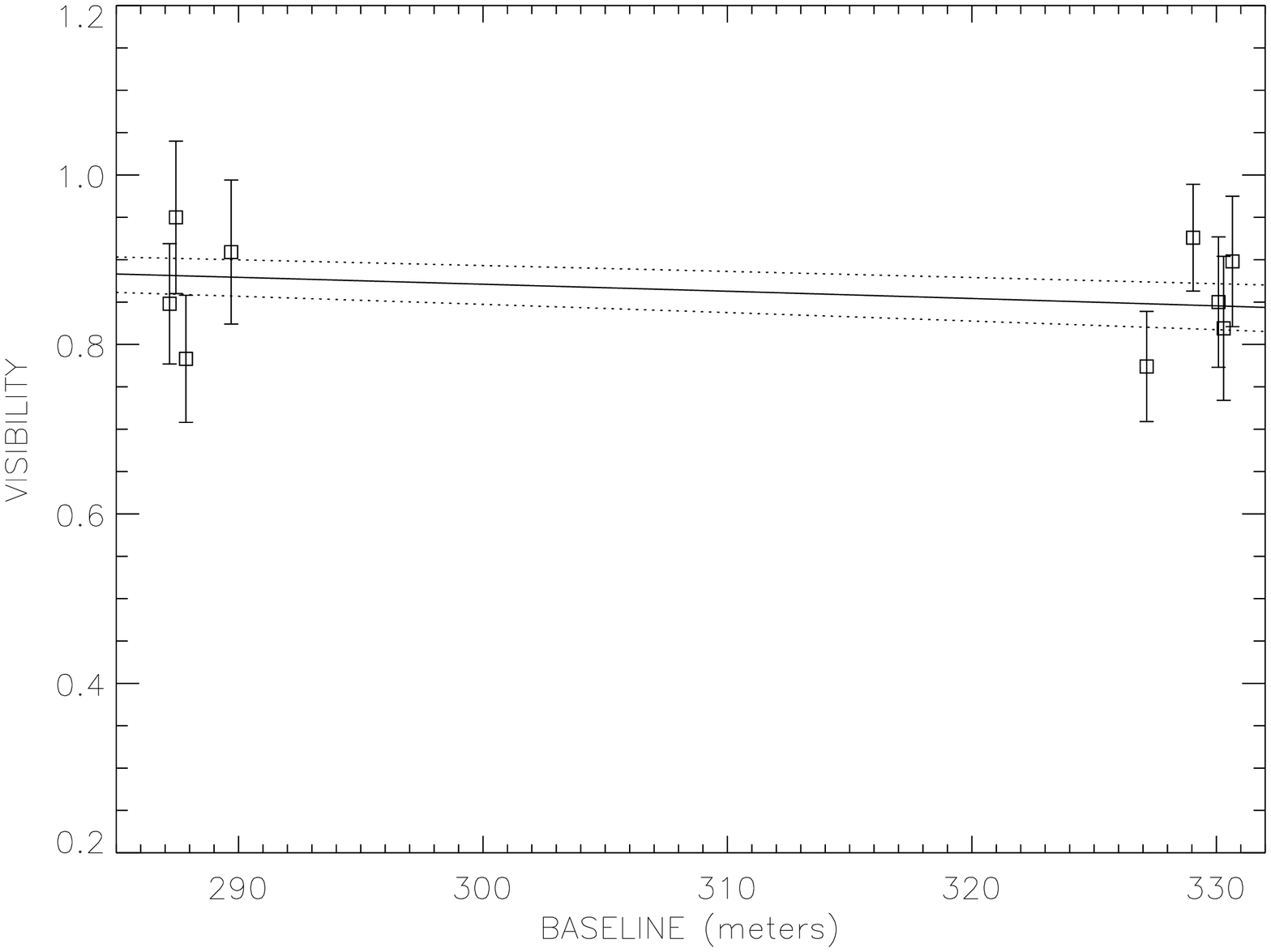}\\
  \caption{HD 196885 limb-darkened disk diameter fit.}
\end{figure}
\clearpage

\begin{figure}[!h]
  \centering \includegraphics[angle=90,width=1.0\textwidth]
  {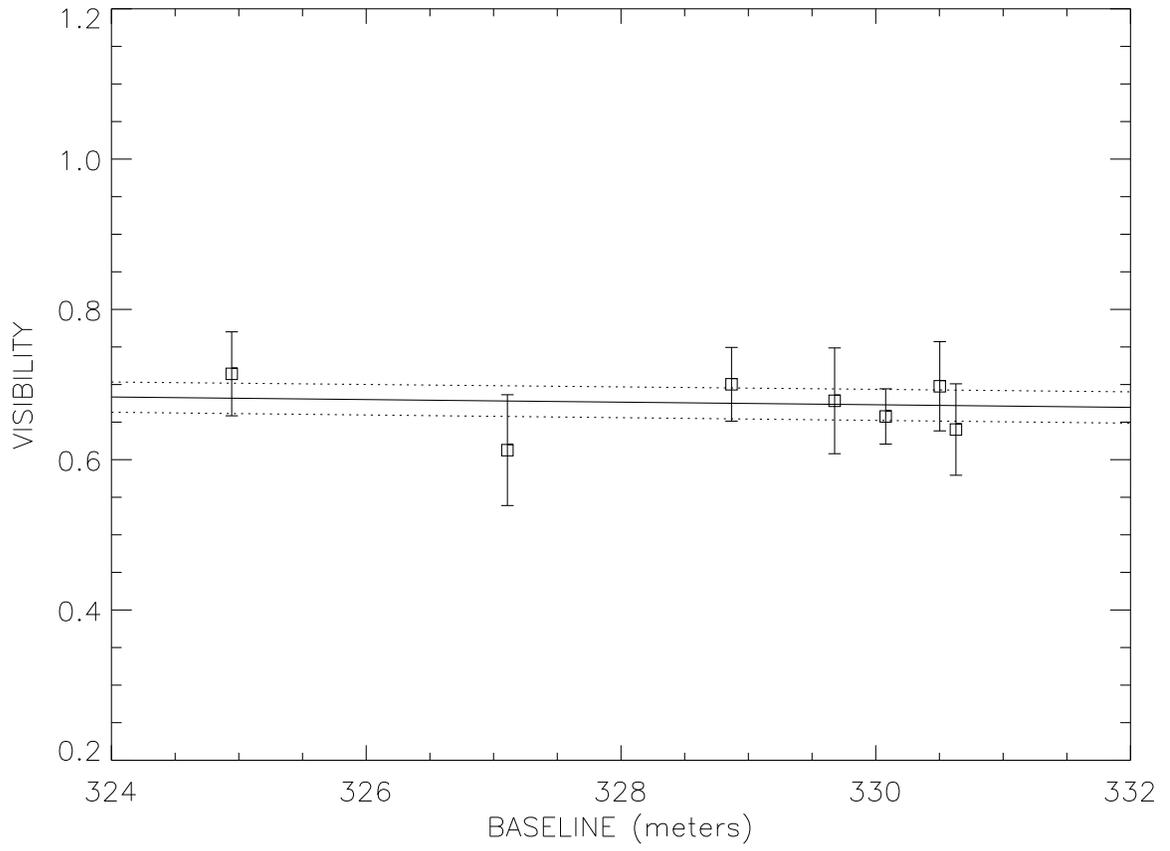}\\
  \caption{HD 217014 limb-darkened disk diameter fit.}
\end{figure}
\clearpage


\begin{figure}[!h]
  \centering \includegraphics[width=1.0\textwidth]
  {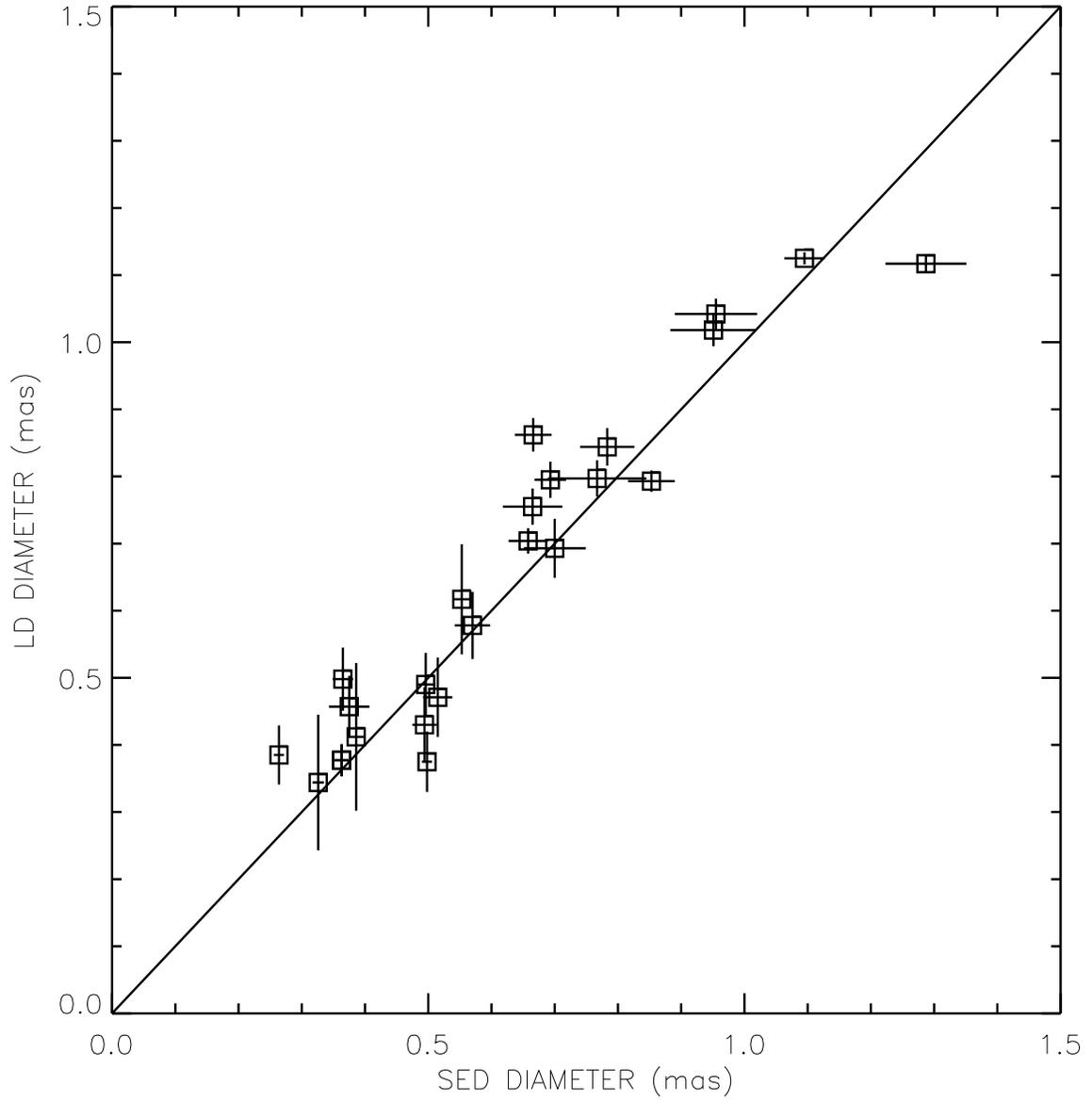}\\
  \caption{A comparison of estimated SED diameters and measured LD diameters. The solid line indicates a 1:1 ratio for the diameters. Note that at $\theta >$ 0.6 mas, the errors for the measured LD diameters become equal to or smaller than the errors from the SED diameter estimates.}
  \label{SED_LDdiams}
\end{figure}

\clearpage


\begin{figure}[!h]
  \centering \includegraphics[angle=90,width=1.0\textwidth]
  {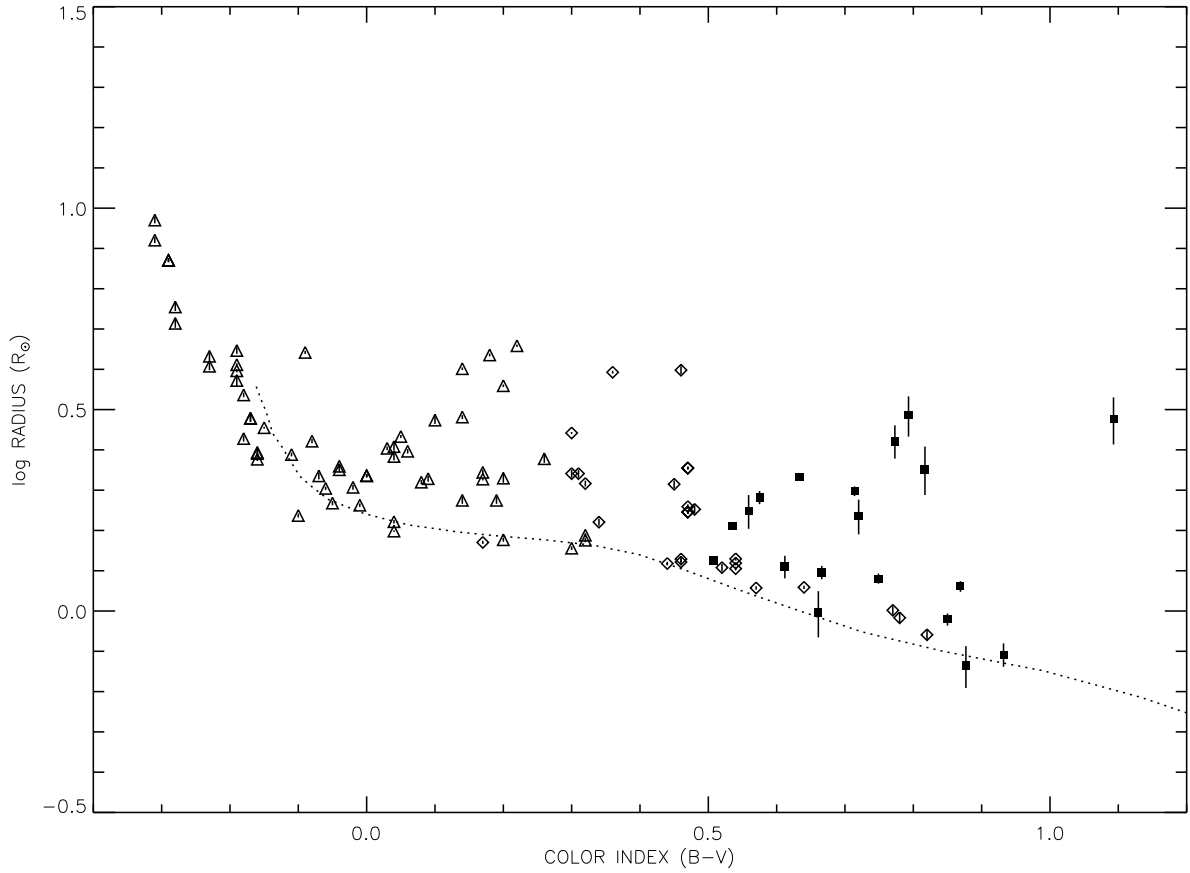}\\
  \caption{Stellar radii: log $R$ vs. unreddened color index ($B-V$). The $\bigtriangleup$s represent O, B, and A-dwarf stars from the Andersen sample \citep{1991A&ARv...3...91A}; $\Diamond$s represent F, G, and K-dwarf stars from the Andersen sample; and the filled $\Box$s represent exoplanet host stars' diameters measured here with errors $<$15\%. The dotted line indicates the ZAMS for stars with masses between 0.15 and 5.0 $M_\odot$ \citep{2000A&AS..141..371G}.}
  \label{rbv_all}
\end{figure}

\clearpage


\begin{figure}[!h]
  \centering \includegraphics[angle=90,width=1.0\textwidth]
  {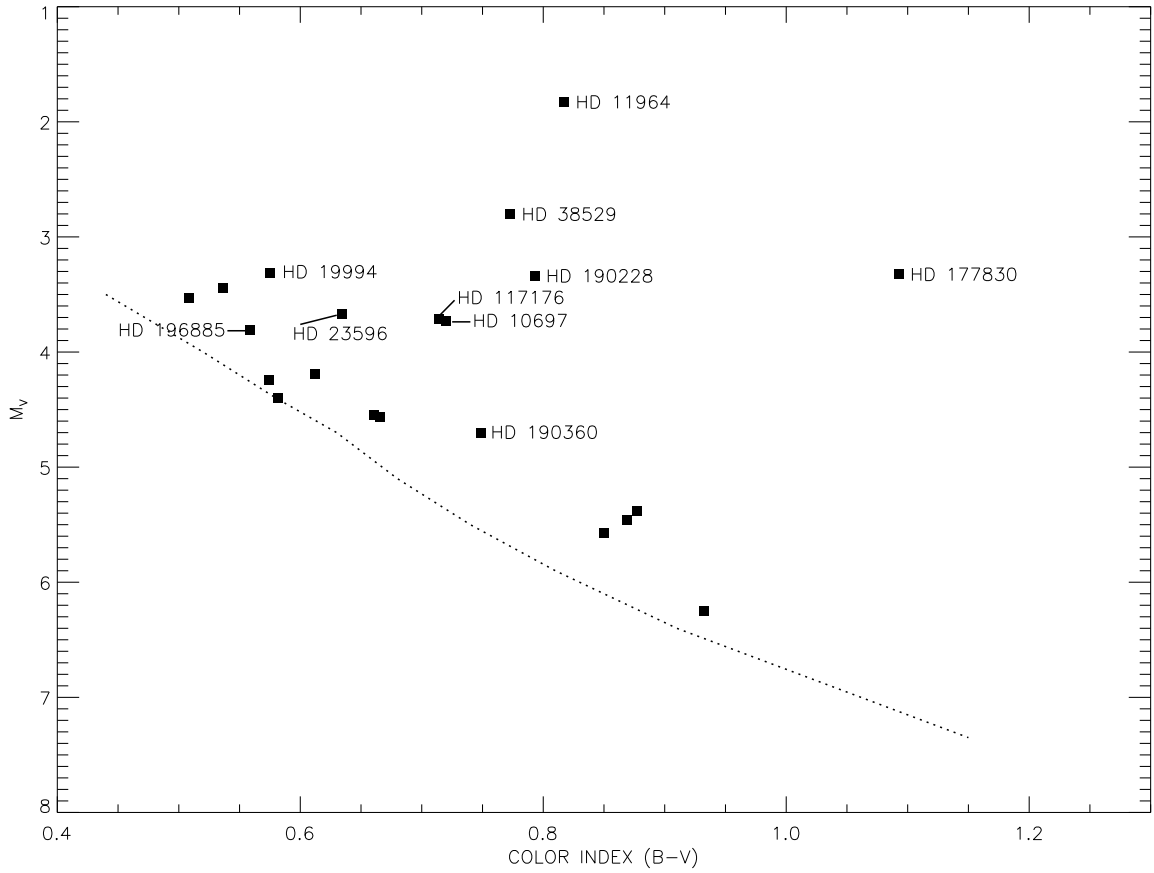}\\
  \caption{Absolute $V$ magnitude vs. color index ($B-V$). The dotted line indicates the ZAMS derived from \citet{2000asqu.book.....C}. The unlabeled points are dwarfs with no problems in their measured radii. HD~10697, HD~11964, HD~38529, HD~177830, and HD~190228 are confirmed subgiants, while HD~19994 and HD~117176 are stars showing signs of post-M-S evolution. See $\S$\ref{dwarfs_subgiants} for details on these stars and on HD~23596, HD~190360, and HD~196885.}
  \label{exsolhr}
\end{figure}

\clearpage

\end{document}